\documentclass[useAMS,fleqn,usenatbib]{mnras}

\usepackage{txfonts}
\usepackage[T1]{fontenc}
\usepackage{longtable}
\usepackage{color}
\usepackage{array}
\usepackage{graphicx}
\usepackage{caption}
\usepackage{url}
\usepackage{aecompl} 
\usepackage{subfig}
\usepackage{float}
\hypersetup{draft}

\newcommand{\RNum}[1]{\uppercase\expandafter{\romannumeral #1\relax}}

\title[Outer Halos of Globular Clusters]{The Outer Envelopes of Globular Clusters. \RNum{2}. NGC 1851,  NGC 5824 and NGC 1261\thanks{This paper includes data gathered with the 6.5 meter Magellan Telescopes located at Las Campanas Observatory, Chile.}}
\author[P. B. Kuzma et al.] {P. B.~Kuzma$^{1}$, G. S.~Da~Costa$^{1}$, A. D.~Mackey$^{1}$\\
$^{1}$Research School of Astronomy \& Astrophysics, Australian National University, Canberra, ACT 2611, Australia; pete.kuzma@anu.edu.au\\
}

\begin{document}

\date{}

\pagerange{\pageref{firstpage}--\pageref{lastpage}} \pubyear{2002}

\maketitle

\label{firstpage}

\begin{abstract}

We present a second set of results from a wide-field photometric survey of the environs of Milky Way globular clusters.  The clusters studied are NGC 1261, NGC 1851 and NGC 5824: all have data from DECam on the Blanco 4m telescope.  NGC 5824 also has data from the Magellan Clay telescope with MegaCam.  We confirm the existence of a large diffuse stellar envelope surrounding NGC 1851 of size at least 240 pc in radius.  The radial density profile of the envelope follows a power-law decline with index $\gamma = -1.5 \pm 0.2$ and the projected shape is slightly elliptical. For NGC 5824 there is no strong detection of a diffuse stellar envelope, but we find the cluster is remarkably extended and is similar in size (at least 230 pc in radius) to the envelope of NGC 1851. A stellar envelope is also revealed around NGC 1261. However, it is notably smaller in size with radius $\sim$105 pc.  The radial density profile of the envelope is also much steeper with $\gamma = -3.8 \pm 0.2$.  We discuss the possible nature of the diffuse stellar envelopes, but are unable to draw definitive conclusions based on the current data. NGC 1851, and potentially NGC 5824, could be stripped dwarf galaxy nuclei, akin to the cases of $\omega$ Cen, M54 and M2.  On the other hand, the different characteristics of the NGC 1261 envelope suggest that it may be the product of dynamical evolution of the cluster.

 \end{abstract}

\begin{keywords}
globular clusters: general; globular clusters: individual (NGC 1261, NGC 1851, NGC 5824, NGC 7089); stars: photometry; Galaxy: stellar content
\end{keywords}

\section{Introduction}

In recent times, a small group of Milky Way Globular Clusters (GCs) have been discovered to possess cluster-like stellar populations beyond their tidal boundaries. The spatial distributions of the extra-tidal populations are found to take two different forms. One takes the shape of two long axisymmetric streams that lead out from the cluster centre, otherwise known as tidal tails \citep[e.g., Palomar 5;][]{2003AJ....126.2385O,2006ApJ...641L..37G}. This feature is suggested to form through the disruption of the parent GC by both internal processes and external tidal forces exerted by the Galaxy \citep[e.g.,][]{2010MNRAS.401..105K}. The other is a diffuse stellar envelope that surrounds the cluster beyond its tidal radius \citep[e.g.,][]{2009AJ....138.1570O,2014MNRAS.445.2971C}. It is currently unclear how diffuse stellar envelopes can form, as there are at least two different theories suggested for their origin.

One is that the diffuse stellar envelope may be a natural product of GC evolution in the Galactic tidal field. Simulations of tidal tail formation do show that a diffuse envelope may form during the disruption process, as stars begin to populate the outermost regions of the cluster before entering the tidal tails \citep[][]{2010MNRAS.401..105K}. Indeed, a number of Galactic GCs do show evidence of stellar envelopes, based on their surface density profiles, that possess similar traits to those seen in the simulations \citep[e.g., see][]{2012MNRAS.419...14C}. Recent work has also shown that observational biases might influence the detection of tidal tails or a stellar envelope in some cases \citep{Balbinot:2017tc}. The other suggested origin regards the stellar envelope as evidence for a cluster having an extra-galactic origin. Studies such as \cite{2009AJ....138.1570O} and \cite{2016MNRAS.461.3639K} have uncovered stellar envelopes that are unlike those expected to be produced during the formation of tidal tails. Further, the envelopes appear to embed massive clusters that have peculiar stellar properties, such as Fe and s-process element variations, which already make them distinct from more "classic" Galactic GCs.

A number of Galactic GCs, whether they have peculiar stellar populations or not, have been linked to an extra-galactic origin. For example, a small handful of GCs are related to the tidal remains of the Sagittarius dwarf galaxy \citep{1994Natur.370..194I,1995AJ....109.2533D,2010ApJ...718.1128L}; whose debris is in the form of a stellar stream that completely wraps around the Milky Way \citep[e.g.,][and references therein]{2003ApJ...599.1082M,2009ApJ...700.1282Y}. Even in M31, our largest galactic neighbour, GCs are seen to inhabit or lie near to large scale stellar structures generated by the tidal disruption of dwarf galaxies \citep[e.g.][]{2002AJ....124.1452F,2010ApJ...717L..11M,2014MNRAS.445L..89M,2014MNRAS.442.2165H}. Therefore, the envelopes may be the remnants of a long since accreted dwarf galaxy, with the rest of the stream remaining undetected.


We are searching for large scale streams in the Galactic halo, using GCs as potential tracers (\citealp{2016MNRAS.461.3639K}, hereafter Paper \RNum{1}). As GCs are seen to lie in or nearby streams in M31, looking at GCs in the Milky Way that hint at an extra-galactic origin with wide field photometry may result in the discovery of streams or some other kind of structure that may be the remnants of the GCs parent dwarf galaxy. Paper \RNum{1} introduced the survey and presented the results of the first cluster examined, NGC 7089 (M2). M2 has peculiar properties that hint at a potential accreted origin (see Paper \RNum{1} and references therein), and wide field imaging revealed that M2 is embedded in a low-mass stellar envelope, detected to a radius of at least $\sim210$ pc, with no signs of a stellar stream or tidal tails. The next set of targets of this survey are the halo GCs NGC 1851, NGC 5824 and NGC 1261.

Many studies of NGC 1851 over the past decade have revealed that the cluster is not a typical Milky Way GC. NGC 1851 is a relatively massive globular cluster, $M_v=-8.33$, located at 12.1 kpc from the sun (Galactocentric distance: 16.6 kpc) \citep[][2010 edition]{1996yCat.7195....0H}. Amongst its peculiarities is a double sub-giant branch in the colour-magnitude diagram \citep{2008ApJ...673..241M}. Other anomalies include a range in C+N+O abundance among the cluster red giants \citep[][and references therein]{2009ApJ...695L..62Y, 2009MNRAS.399..934V, 2015MNRAS.446.3319Y}, and star-to-star variations in [Fe/H] and s-process elements \citep{2008ApJ...672L..29Y,2010ApJ...722L...1C,2012A&A...544A..12G}. The two sub-giant branch populations correlate with the observed abundance variations, as the brighter sub-giant branch is metal-poor and under-abundant in s-process elements compared to the fainter sub-giant population. Lastly, \citet{2009AJ....138.1570O} uncovered the existence of a stellar halo surrounding the cluster, that is at least 500pc in diameter. NGC 1851 halo stars have been identified through radial velocities \citep[e.g.,][]{2014MNRAS.442.3044M,2015MNRAS.453..531N} and have been found to exhibit the same s-process abundance patterns as the bright sub-giant branch stars within the cluster, confirming that the envelope is directly related to NGC 1851 \citep{2014MNRAS.442.3044M}.

The origin of the halo embedding NGC 1851 has been the source of much speculation. \cite{2009AJ....138.1570O} discussed its formation in terms of originating from the cluster. This scenario is suggested as unlikely due to the lack of evidence for tidal tails in their observations. The second suggestion made by the authors is that the halo is evidence for NGC 1851 being the core a dwarf galaxy that has been accreted by the Milky Way. As discussed above, NGC 1851 has anomalies in common with a group of Milky Way GCs that all have been linked to the remains of, or belonging to, an accreted dwarf galaxy (e.g., M54, $\omega$ Cen, M2). \cite{2012MNRAS.419.2063B} modelled the halo/cluster system of NGC 1851, exploring different potential formation scenarios. They found NGC 1851 could have formed in the central regions of a dwarf galaxy as a product of GC-merging at the center, or be the actual nucleus itself. Regardless of which scenario, the models predict what we currently observe as regards the envelope of NGC 1851.

NGC 5824 ($M_v = -8.85$, heliocentric distance: 32.1 kpc), the third target of this study, shows some of the same characteristics as NGC 1851 and M2. It may have a [Fe/H] abundance spread: \cite{2014MNRAS.438.3507D} investigated the alluring result of a possible Fe spread in NGC 5824 by \cite{2012A&A...540A..27S}, inferred from medium-resolution spectroscopy of the Ca \RNum{2} triplet. \cite{2014MNRAS.438.3507D} found a $\sim0.3$ dex range in [Fe/H] across their sample of red giant branch stars. However, a recent analysis based on high dispersion spectra of NGC 5824 red giants by \citet{2016MNRAS.455.2417R} excluded the existence of a Fe-spread of this size. Nevertheless, \citet{2016MNRAS.455.2417R} also showed that NGC 5824 has a large [Mg/Fe] variation and that one star in their sample has significantly higher s-process element abundances than the others. 

The cluster also satisfies the criteria of possessing cluster stars that exist beyond the tidal radius. \cite{1995AJ....109.2553G} included NGC 5824 in their search for GCs with tidal tails. Their star counts revealed a radial profile that deviates from a King model near the tidal radius and continues to drop off at rate well described by a power law, $\gamma = -2.2 \pm 0.1$ \citep{2014MNRAS.438.3507D}. More recently, \cite{2012MNRAS.419...14C} presented a density profile of NGC 5824 in their sample of globular clusters observed with the ESO Wide Field Camera and also found a profile that is described by a power law ($\gamma = -2.62$). \citet{2009ApJ...700L..61N} presented a possible connection of NGC 5824 with the Cetus Polar stream, based on orbit calculations for the stream. However, later follow up analysis of the spatial distribution of cluster stars by \cite{2014MNRAS.445.2971C} did not reveal any extended structure.

The fourth and final cluster in this paper is NGC 1261. It is slightly further away than NGC 1851, 16.3 kpc, and it is a little fainter, $M_v=-7.80$. It has no known measured internal Fe-dispersion, though there is tantalising suggestion of a potentially small population with a heavy element dispersion in the chromosome maps of \citet[][]{2016MNRAS.tmp.1516M}. This metal-poor cluster was part of the \citet{2014MNRAS.445.2971C} sample, and these authors claim the potential existence of stellar population beyond 1.5 times the tidal radius determined in \citet{2012MNRAS.419...14C}. The authors do not suggest whether this extra population is in tidal tails, or is an envelope, or a sign of a larger stellar stream. They lament their lack of sufficient spatial coverage to investigate more fully the origin. Quite recently, \cite{2016ApJ...820...58B} present a new discovery of a stellar stream, the Phoenix stream, in the Dark Energy Survey's first year data that lies in the direction of NGC 1261, presenting a promising connection at first. However, the authors comment that there is no clear connection between NGC 1261 and the Phoenix stream. The authors also dismiss a potential connection between the stream, NGC 1261 and the newly discovered Eridanus-Phoenix over-density, a stellar structure that is same direction and at a similar heliocentric distance \citep{2016ApJ...817..135L} as the cluster.

This paper presents the results of wide-field imaging, using MegaCam and the Dark Energy Camera (hereafter DECam), of NGC 1851, NGC 5824 and NGC 1261. A brief list of parameters for these clusters is presented in Table \ref{tab:clusters}. The observations and the reduction techniques involved are discussed in the following section, while in section 3 we present our results. We analyse our findings in section 4 and discuss the results in section 5. Our concluding comments are presented in section 6.

\begin{figure*}
  \begin{center}  
    \includegraphics[width=0.90\textwidth]{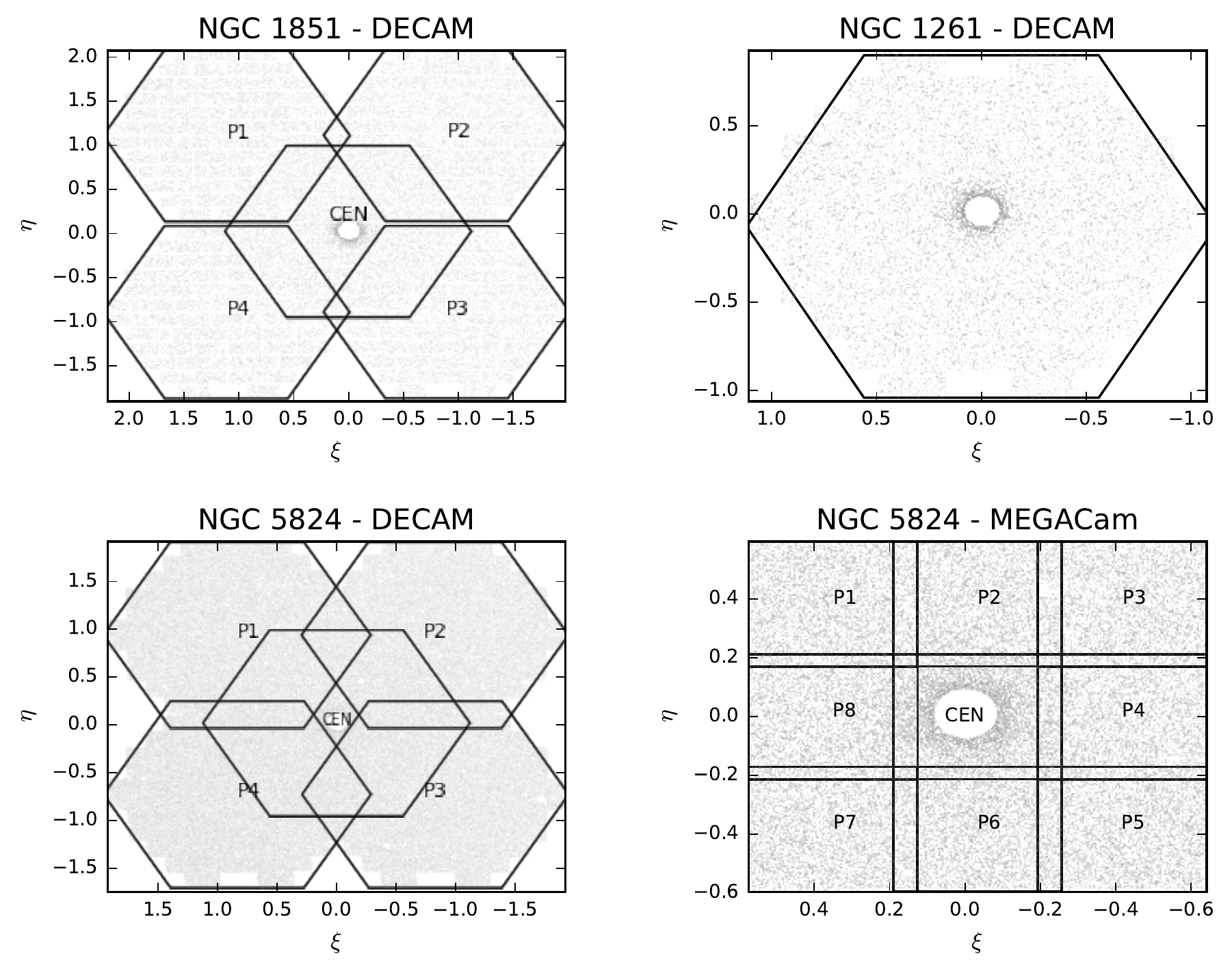}
  \end{center}
\caption{Locations of the fields of view in this study, plotted on the gnomonic projection ($\xi, \eta$) in degrees. Also plotted are the stars extracted from the images. Due to crowding, we have excluded the inner regions of the clusters. North is up and East is to the left. Regions excluded: NGC 1851 - 6 arcmin radius, NGC 1261 - 5 arcmin, NGC 5824 (both cameras) - 5 arcmin.}
\label{fig:FIELD}
\end{figure*}
\begin{table*}

\begin{minipage}{160mm}
\begin{center}
\caption{A brief list of parameters for the clusters in this paper.}

\label{tab:clusters}
\begin{tabular}{@{}cccccccccc}

\hline \hline
Cluster&R. A.$^1$&Dec.$^1$&$l$&$b$&[Fe/H]$^2$&\multicolumn{2}{c}{Distances$^{3}$}&$M_{v}$$^{3}$\\
&\multicolumn{2}{c}{(J2000)}&\multicolumn{2}{c}{(deg)} &(dex)&Solar (kpc)& Galactocentric (kpc)&(mag)\\
\hline
NGC 1261&03:12:16&-55:12:58&270.54&-52.12&-1.27&16.3&18.1&-7.80\\ 
NGC 1851&05:14:07&-40:02:48&244.51&-35.03&-1.18&12.1&16.6&-8.33\\
NGC 5824&15:03:59&-33:04:06&332.56&22.07&-1.94&32.1&25.9&-8.85\\

\hline

\end{tabular}
\\References: $^{1}$ \citet{2010AJ....140.1830G}, $^{2}$ \citet{2009A&A...508..695C}, $^{3}$ \citet{1996yCat.7195....0H} (2010 edition)
\end{center}

\end{minipage}
\end{table*}

\section{Observations and Data reduction}
\subsection{Observations}
The observations of NGC 1851 and NGC 5824 were taken with the Dark Energy Camera (DECam; \citealt{2015AJ....150..150F}) on the 4-m Blanco telescope at Cerro Tololo Inter-American Observatory. DECam, a mosaic imager, has 62 CCDs arranged in a near circular pattern with each CCD containing 2048 x 4096 pixels. Each cluster was observed as sets of three dithered exposures in the \textit{g} and \textit{i} filters with five separate pointings.  The pointings are a central one and four locations along the diagonals that define an x-shape (see Fig. \ref{fig:FIELD}). The observations for NGC 1851 were taken on February-17, 2013, while the observations for NGC 5824 were performed on the August-13, 2013, and February-26, 2014. The exposure time for NGC 1851 was 200 sec for both filters, and saw mostly consistent seeing: \textit{g} seeing was between 0.92\arcsec - 1.02\arcsec  and 0.83\arcsec - 0.96\arcsec  for the \textit{i} images. We adopted the seeing as the full-width at half-maximum (FWHM) of the stellar images. NGC 5824 had a slightly longer exposure time, 360 seconds per exposure for both filters, and more variable seeing. The \textit{g} observations had seeing between 1.25\arcsec - 1.67\arcsec,  and the \textit{i} observations had a range of 1.13\arcsec - 1.25\arcsec. A second set of observations for the central pointing of NGC 5824 was obtained with significantly shorter exposure time, 10 seconds, on March-06, 2016, in both filters for the purposes of photometric calibration.

The final set of DECam observations, for NGC 1261, was obtained on February-28, 2014. Unlike NGC 1851 and NGC 5824, NGC 1261 had only one field taken, dithered three times, with the cluster in the center. Exposures were shorter for \textit{g} (250s) with rather consistent seeing conditions (1.03\arcsec - 1.07\arcsec) than for \textit{i} (360s), which had slightly variable seeing (0.80\arcsec - 0.92\arcsec). All the observations were processed through the DECam community pipeline reduction pipeline\footnote{\url{http://www.ctio.noao.edu/noao/content/dark-energy-camera-decam}} \citep{2014ASPC..485..379V}. 
NGC 5824 also had a set of observations obtained with MegaCam \citep{2015PASP..127..366M} on the 6.5-m Magellen Clay telescope at Las Campanas Observatory, undertaken on June-14 and June-15, 2013. Nine pointings, dithered three times, were observed in a three by three grid, with the central observation containing NGC 5824 itself (Fig. \ref{fig:FIELD}, bottom right). The \textit{g} observations were exposed for 90 seconds, and saw seeing variability between 0.96\arcsec - 1.33\arcsec. The \textit{i} observations used a 300 second exposure time, and had seeing in the range 0.77\arcsec - 1.15\arcsec. The observations were processed and reduced at the Harvard-Smithsonian Center for Astrophysics with the MegaCam pipeline \footnote{\url{http://hopper.si.edu/wiki/piper/Megacam+Data+Reduction}} \citep[see][]{2015PASP..127..366M}.

The full details of the DECam and MegaCam observations are given in the accompanying on-line material.  In that material we provide for each cluster the field names and field-centre positions, the number of exposures, the exposure times and the seeing for both the \textit{g} and \textit{i} images.

\begin{figure*}
  \begin{center}  
    \includegraphics{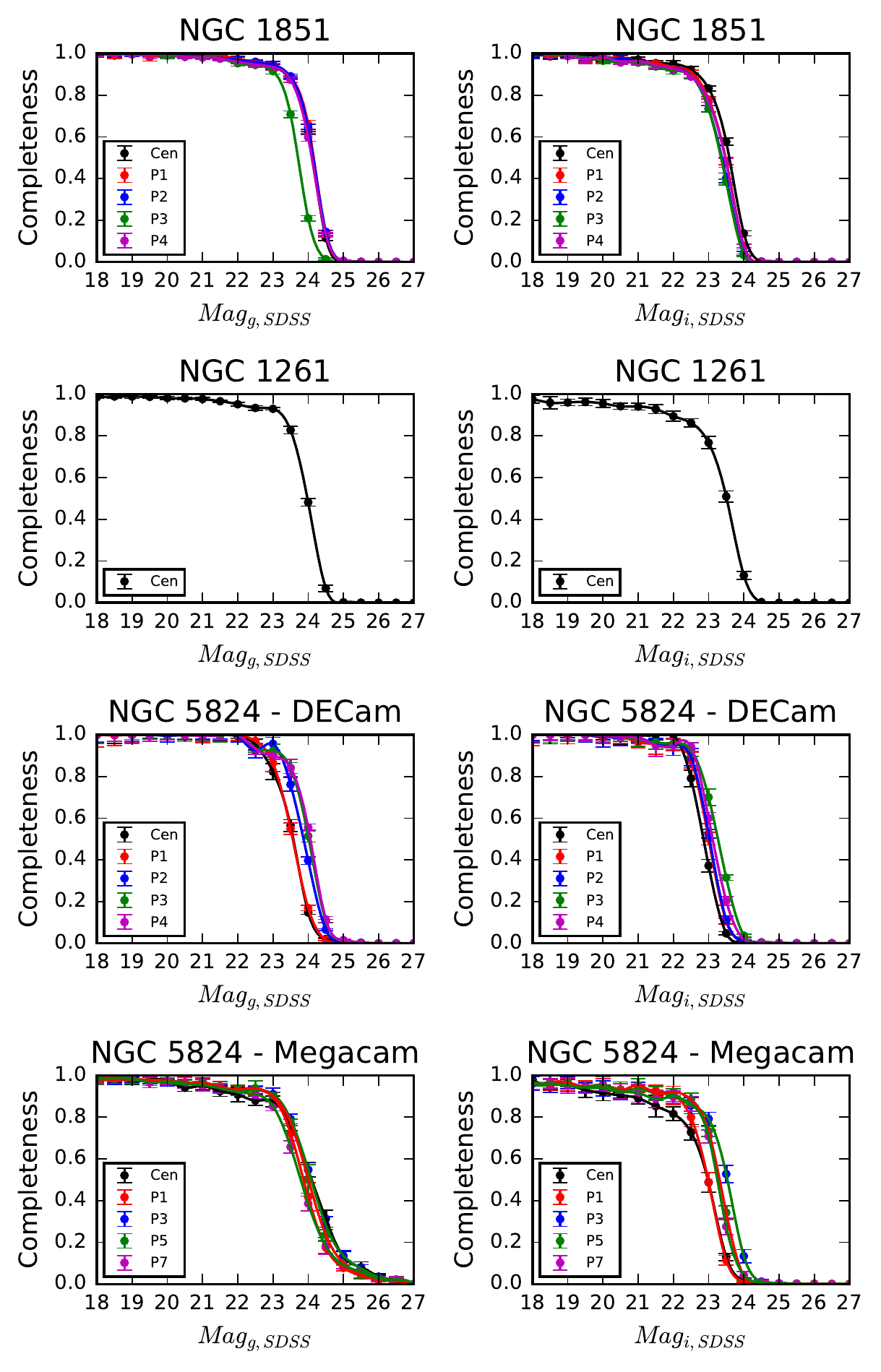}
  \end{center}
\caption{Completeness for our set of observations, along with a best fit cubic spline. For clarity, we display only a subset of completeness curves for the MegaCam observations of NGC 5824.}
\label{fig:CMPL}
\end{figure*}

\begin{figure*}
  \begin{center}  
    \includegraphics[width=0.90\textwidth]{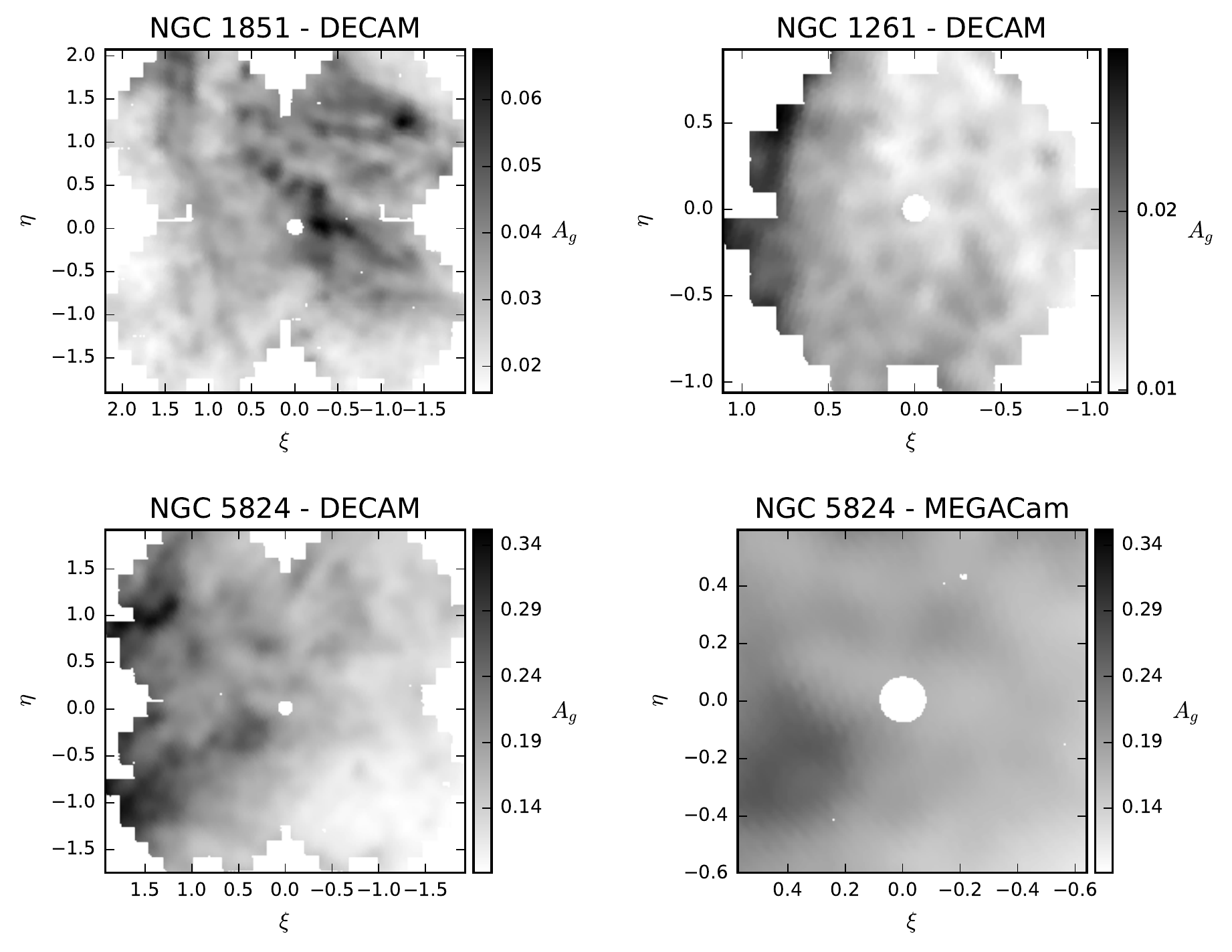}
  \end{center}
\caption{The $A_g$ extinction maps for across the fields \citep{1998ApJ...500..525S,2011ApJ...737..103S}. Note the strong variable reddening across the DECam imaging of NGC 5824. }
\label{fig:DUST}
\end{figure*}

\subsection{Photometry}
The entire pipeline that takes the reduced photometric images to a workable photometric catalog is discussed in detail in Paper \RNum{1}; here we will outline the major components. The software package Source Extractor\footnote{\url{https://www.astromatic.net/software/sextractor}} \citep[SExtractor;][]{1996A&AS..117..393B} was used to perform aperture photometry on the non-stacked, individual photometric images. Each resulting image catalog was cleaned of spurious and non-stellar objects through a combination of SExtractor output flags and magnitude differences between different aperture sizes. The cleaned \textit{g} and \textit{i} catalogs for each image were cross-matched using the freely available software, Stilts \citep{2006ASPC..351..666T}. 

Once each image had their \textit{g} and \textit{i} catalogs generated, all catalogs belonging to the same field were combined to create a single stellar catalog for each field. Before this occurred, however, the catalogs were adjusted to be placed on the same photometric scale. The image with the deepest photometry was designated the master image, and the remaining two images were adjusted to the master by determining the median photometric difference with stars observed in other two images. The single stellar catalogs for each field were then calibrated to a master field using the same techniques: for DECam and MegaCam, that field was the central pointing (CEN). NGC 1261 only had one pointing, therefore no cross-field calibrations were necessary. 

Finally, the complete catalogs were transferred to a known photometric scale from the raw instrumental magnitudes. For NGC 1851 and NGC 1261, we cross matched stars with a catalog belonging to the Dark Energy Survey (DES; \citealt{2005astro.ph.10346T}), which is calibrated to the Sloan photometric scale and is presented in \citet{2015ApJ...805..130K}. Table \ref{tab:SDSS-DC} displays our photometric calibration corrections (zero point and colour terms) for those two clusters. As for NGC 5824, the short exposure set of DECam observations for the central pointing (CEN) was used to calibrate the instrument photometry to APASS \citep{2009AAS...21440702H,2015AAS...22533616H}, which is also on the Sloan photometric scale. Using stars in common again, we calibrated the shallower photometry to the APASS catalog. From there, the deeper imaging was calibrated to the shallow DECam imaging. Finally, the MegaCam catalog was then scaled to APASS through the stars that were mutually observed in DECam and MegaCam. Table \ref{tab:SDSS-MC} lists the zero points and colour terms of the NGC 5824 photometric corrections. 

\begin{table}
\begin{center}
\caption{The parameters used to calibrate our instrumental photometry to the SDSS system.}
\label{tab:SDSS-DC}
\begin{tabular}{@{}cccc}
\hline \hline
Cluster&Filter &\multicolumn{2}{c}{Calibration}\\
&&Zero Point & Colour Coeff.\\
\hline
NGC 1851&\textit{g}&$30.755\pm0.002$&$-0.016\pm0.001$\\
&\textit{i}&$31.753\pm0.002$&$0.002\pm0.001$\\
NGC 1261&\textit{g}&$30.829\pm0.002$&$-0.015\pm0.001$\\
&\textit{i}&$31.289\pm0.002$&$0.001\pm0.001$\\
\hline
\end{tabular}
\end{center}
\end{table}

\begin{table}
\begin{center}
\caption{The parameters used to calibrate our NGC 5824 instrumental photometry to the APASS system.}
\label{tab:SDSS-MC}
\begin{tabular}{@{}cccc}
\hline \hline
Camera&Filter &\multicolumn{2}{c}{Calibration}\\
&&Zero Point & Colour Coeff.\\
\hline
MegaCam $\to$ APASS &\textit{g}&$30.987\pm0.006$&$-0.174\pm0.003$\\
&\textit{i}&$31.610\pm0.002$&$0.012\pm0.001$\\
DECam $\to$ APASS&\textit{g}&$31.321\pm0.006$&$-0.028\pm0.006$\\
&\textit{i}&$31.285\pm0.0008$&$0.039\pm0.007$\\
\hline
\end{tabular}
\end{center}
\end{table}

\subsection{Artificial Star Tests}
To test the recovery rate of the pipeline and to explore the completeness across the mosaics, we have performed artificial star tests on all the observations. 10000 stars and 2000 stars were inserted for DECam and MegaCam respectively. The artificial stars were the brightest at 18$^{th}$ magnitude, and increased in frequency towards the faint limit of 27$^{th}$ magnitude. Stars were deemed `recovered' if they passed through the same pipeline described in the previous section. This process was repeated 10 times, to create a sizeable sample of stars for each camera. For DECam, this amounted to 100000 stars per field, and 20000 stars for MegaCam.

To ensure variable completeness does not influence the results we limited all the photometry to the 90\% completeness magnitude for the field with the shallowest photometry (see Fig. \ref{fig:CMPL}). With respect to NGC 1851, this photometric limit was set in \textit{g} by P3:  \textit{g}$=23.0$, while in \textit{i} each field appeared consistent with each other and the limit is \textit{i}$=22.3$. The corresponding 90\% completeness for NGC 1261 are 23.4 in \textit{g} and 22.6 in \textit{i}.  As for NGC 5824, the cluster is considerably more distant than the others. Therefore, to maximise coverage of the main sequence, we adopted a 50$\%$ cut rather than 90$\%$. This allowed increased photometric depth without the photometric uncertainties becoming too large. For DECam, this corresponded to 23.4 in \textit{g} and 22.7 in \textit{i}. MegaCam's corresponding magnitudes were 23.7 and 22.8 for \textit{g} and \textit{i} respectively. We also explored the completeness as a function of radius from the cluster center to examine the radius that relative completeness becomes affected by crowding. When completeness became significantly affected, we excluded all detections within the corresponding radius. This radius was 6$\arcmin$ for NGC 1851 and 5$\arcmin$ for NGC 1261 and NGC 5824. However, we affirm that our analysis is predominately undertaken outside the limiting radii of our clusters. At these distances, the completeness curves in our fields do not vary significantly.

\subsection{Extinction}
Fig. \ref{fig:DUST} displays the extinction in the field of view for our observations from the dust maps presented by \cite{1998ApJ...500..525S}. NGC5824 is affected by a severe amount of variable reddening along our line of sight, $0.09 \leq A_g \leq 0.35$. This is seen as well for NGC1851 and NGC 1261 ($0.01 \leq A_g \leq 0.07$ and $0.01 \leq A_g \leq 0.03$, respectively), though not at the same level of severity. If not taken into consideration, much like relative completeness across the fields of view, the variable reddening may lead to apparent low surface brightness features that are a product of reddening and not real. After the photometric cuts due to completeness, we corrected each star individually for reddening (now denoted $g_0$ and $i_0$) and performed another photometric cut based on the level of reddening. We found the region in the fields that had the most severe level of extinction and removed all stars from the catalogs that were fainter than the limiting magnitude of that region. This meant all stars fainter than 22.7 in $g_0$ and 22.2 in $i_0$ were removed for NGC 1851. As for NGC 5824, the new DECam limits were 22.9 and 22.1 in $g_0$ and $i_0$ respectively, and 22.9 ($g_0$) and 22.4 ($i_0$) for MegaCam. Finally, the new photometric limits for NGC 1261 are 23.2 in $g_0$ and 22.5 in $i_0$.

\subsection{Complete Catalog}
The complete DECam photometric catalogs of our three clusters are displayed in the colour-magnitude diagrams (CMD) in Fig. \ref{fig:CMD}. The main sequence and the main sequence turn-off are clearly visible in all clusters, as well as a noticeably populated blue horizontal branch for NGC 5824. We have removed the field dwarfs from any subsequent analysis as they are clearly contaminants not related to the cluster population. This means we removed stars with $(g_0-i_0)>1.6$ in the catalogs belonging to NGC 1851 and NGC 1261, and $(g_0-i_0)> 0.8$ for NGC 5824. As is evident in Fig. \ref{fig:CMD}, the red giant branch in all three cluster CMDs is essentially undetectable against the field contamination, while the main sequences are readily visible.  Consequently, we decided to use only stars fainter than approximately the level of the main sequence turnoff in the analysis.  The cutoff magnitudes were $i_{0} = 18.5, 19.0$ and 20.5 for NGC 1851, NGC 1261 and NGC 5824, respectively. Fig. \ref{fig:CMD_MC} displays the CMD for the NGC 5824 MegaCam catalog. This catalog underwent the same photometric cuts as the DECam catalogs.

\section{Results}
In this section we present our results using the techniques previously discussed in depth in Paper \RNum{1}, unless otherwise specified. As the techniques are described briefly here, we refer any reader requiring more information about the techniques to Paper \RNum{1}. The techniques are common between both cameras and clusters: any differences or cluster specific techniques are mentioned and any new techniques introduced in this analysis will be explicitly stated. 

\begin{table*}
\begin{minipage}{160mm}
\begin{center}
\caption{Fitted structural parameters from the LIMEPY surface brightness models.}

\label{tab:LIMEPY}
\begin{tabular}{@{}ccccccc}

\hline \hline
Cluster& Model$^{1}$ & $W$ & $r_c$ & $r_t$ & $r_h$&$c$\\
(1)&(2)&(3)&(4)&(5)&(6)&(7)\\
\hline
M2&K&$7.052\pm0.053$&$0.333\pm0.003$&$11.641\pm0.287$&$1.340\pm0.020$&$1.543\pm0.026$\\
&W&$6.176\pm0.014$&$0.437\pm0.003$&$33.036\pm0.385$&$1.278\pm0.005$&$1.879\pm0.014$\\
NGC1261&K&$5.856\pm0.090$&$0.352\pm0.003$&$5.836\pm0.309$&$0.890\pm0.035$&$1.220\pm0.054$\\
&W&$5.172\pm0.173$&$0.416\pm0.016$&$12.651\pm1.019$&$0.882\pm0.028$&$1.483\pm0.089$\\
NGC1851&K&$7.988\pm0.064$&$0.095\pm0.002$&$6.397\pm0.266$&$0.669\pm0.028$&$1.828\pm0.047$\\
&W&$7.207\pm0.013$&$0.118\pm0.003$&$41.541\pm0.696$&$0.636\pm0.008$&$2.547\pm0.030$\\
NGC5824&K&$8.204\pm0.124$&$0.080\pm0.009$&$6.277\pm0.693$&$0.662\pm0.074$&$1.895\pm0.158$\\
&W&$7.606\pm0.011$&$0.076\pm0.002$&$57.169\pm0.694$&$0.612\pm0.074$&$2.876\pm0.158$\\
\hline

\end{tabular}
\\(1)Cluster Name, (2) Type of model fit: K - King (1966), W - Wilson (1975), (3) Dimensionless central potential, (4) core radius in arcmin, (5) truncation radius in arcmin, (6) half-mass radius in arcmin, (7) central concentration, $c \equiv \log{r_t/r_c}$

\end{center}

\end{minipage}
\end{table*}

\subsection{Field Identification and Subtraction}\label{chap:BIS}
\begin{figure*}
  \begin{center}  
    \includegraphics[width=\textwidth]{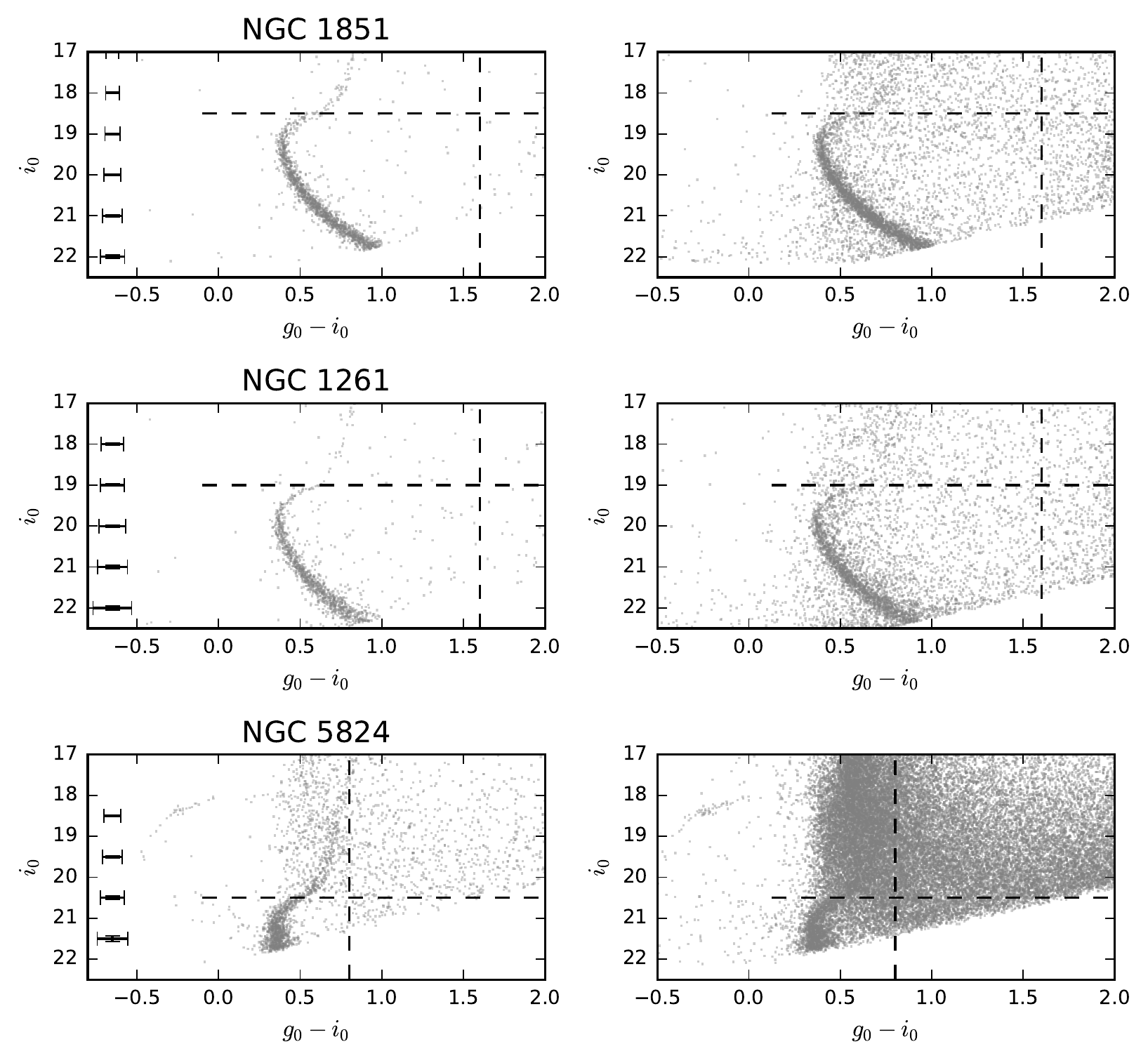}
  \end{center}
\caption{Colour-magnitude diagrams (CMD) of our clusters from our DECam images. Left col: All stars detected at distances from the cluster center $6\arcmin < r <10\arcmin$ for NGC 1851, $5\arcmin < r < 10\arcmin$ for NGC 1261 and NGC 5824, are plotted to show the cluster main sequences. Right col: All stars at $6\arcmin < r < 40\arcmin$ for NGC 1851 and $5\arcmin < r < 40\arcmin$ for NGC 1261 and NGC 5824. The horizontal and vertical dashed lines indicate the photometric cuts undertaken in this paper.} \label{fig:CMD}
\end{figure*}

\begin{figure}
  \begin{center}  
    \includegraphics[width=\columnwidth]{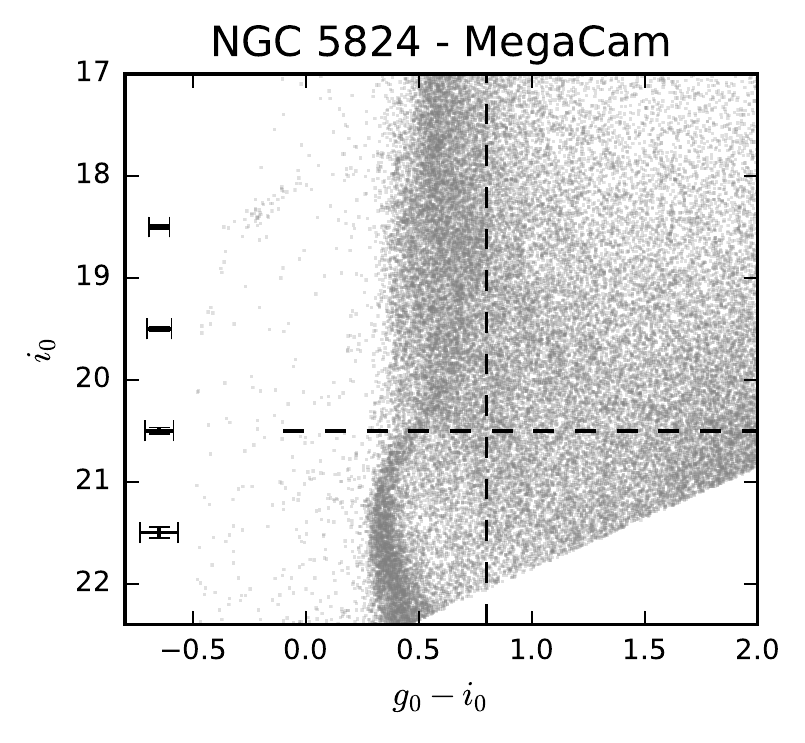}
  \end{center}
\caption{CMD of NGC 5824 from our MegaCam observations.  Stars shown lie within 5 and 40 arcmin of the cluster center. Horizontal and vertical dashed lines indicate the photometric cuts as in Fig. \ref{fig:CMD}.} \label{fig:CMD_MC}
\end{figure}

\begin{figure*}
  \begin{center}  
    \includegraphics[width=0.8\textwidth]{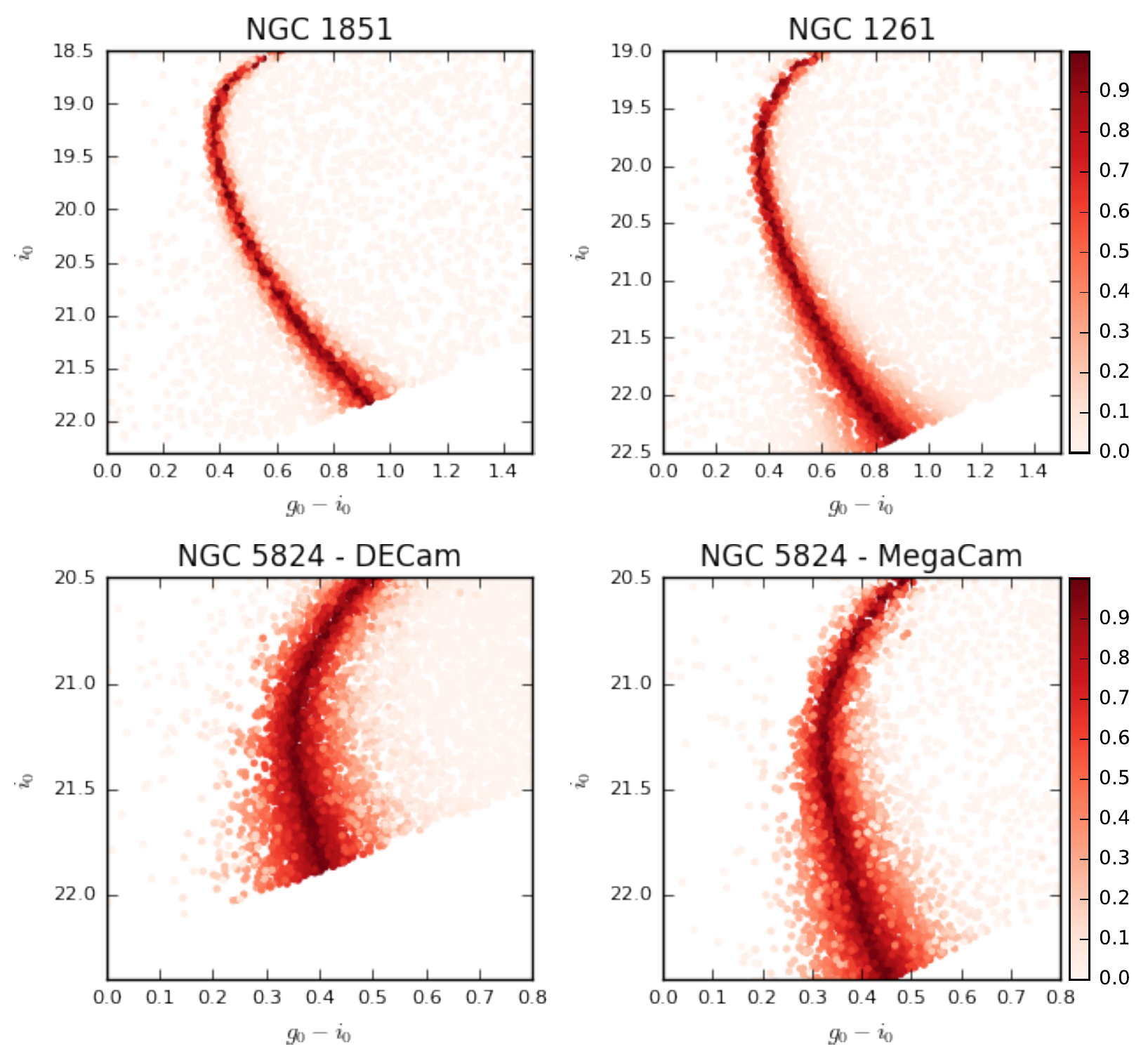}
  \end{center}
\caption{The isochrone-based weighting scheme for the globular clusters in this study for stars in the radial regions depicted in the right column of Fig. \ref{fig:CMD}. Starting top right moving clockwise: NGC 1851, NGC 1261, NGC 5824 MegaCam and NGC 5824 DECam. The points have been colour-coded to show their given weight. }
\label{fig:ISO}
\end{figure*}

For all these clusters, as shown in Figs. \ref{fig:CMD} and \ref{fig:CMD_MC}, the main sequence is pronounced and it is these stars that we will use to construct the radial density profiles. The main sequences were fit with isochrones from the Dartmouth Stellar Evolution Database\footnote{\url{http://stellar.dartmouth.edu/~models/index.html}} \citep{2008ApJS..178...89D}. The isochrones employed were NGC 1851: age $ = 10.5$ Gyr, [Fe/H] $= -1.18$, [$\alpha$/Fe] $= +0.4$; NGC 1261: age $= 10.5$ Gyr, [Fe/H] $= -1.27$, [$\alpha$/Fe] $ = +0.4$; NGC 5824: age $= 10.2$ Gyr, [Fe/H] $= -1.91$, [$\alpha$/Fe] $= +0.4$. While the isochrone ages are slightly younger than the usually accepted ages of globular clusters, they provide the best description of the main sequences and main sequence turn-offs in the CMDs.  Using the isochrones, we can begin removing unwanted foreground and background stars that contaminate our stellar catalogs. To do this, we used an isochrone-weighting scheme that is the same technique as presented in paper I. For each cluster, stars were assigned a weight value {\it w} with values between 0 and 1, that was based on their $g_0-i_0$ colour difference with respect to the isochrone at the $i_0$ of the star. The weight is a Gaussian probability of the observed colour difference given the observational error in colour at the star's magnitude \citep{2015ApJ...804..134R}.  For example, a star on the isochrone has weight 1.0 and a $\pm1\sigma$ colour error deviate has weight 0.61. For a pure cluster star population the distribution of weight values is strongly peaked towards higher values, while for a pure foreground population the distribution of {\it w} values is approximately uniform.  The CMD diagrams for the clusters, coded by the {\it w} values are shown in Fig. \ref{fig:ISO}.

For each cluster the distribution of {\it w} values was evaluated to establish the appropriate {\it separation weight $w_{sp}$} used to distinguish `cluster stars' ({\it w $\geq$ $w_{sp}$}) from `field' stars ({\it w < $w_{sp}$}).  A large {\it $w_{sp}$} value imposes a narrow window on the CMD about the fitted isochrone maximising the relative number of cluster stars but reducing the total sample.  A small value of {\it $w_{sp}$} generates a larger sample but necessarily increases the number of field stars included.  Given the different levels of field contamination, the distribution of the {\it w} values for each cluster are different, and, as a consequence, the adopted value of the separation weight for each cluster varied.  We adopted the separation weight as the value that reproduced the appearance of the cluster main sequences shown, for example, in the left panels of Fig \ref{fig:CMD}.  The {\it $w_{sp}$}values adopted were 0.2 for NGC 1851, 0.3 and 0.15 for NGC 5824 (DECam and MegaCam, respectively), and 0.4 for NGC 1261.  Admittedly the choice of the {\it $w_{sp}$}values used is a judgement call, but trials conducted with varying the {\it $w_{sp}$}values used indicated that changes of order 0.05 do not alter the results.  Substantially larger values reduce the sample sizes making it difficult to detect cluster stars at large radii, while substantially small values increase the field contamination which swamps the cluster-star signal at large radii.

We note, however, that not all the stars above the adopted separation weights are necessarily cluster member stars: field stars that happen to fall within the cluster-star window in the CMD will remain. We address this issue in the following analysis.


The successful removal of the field contamination is important for the validity of any extra tidal detections.  To remove the field star contamination within our `cluster-stars' sample, we first calculated azimuthally-averaged radial profiles from the `cluster stars' sample using a gnomonic coordinate projection. The density profiles were computed using the same method as described in Paper I: densities were calculated for a series of concentric circular annuli of increasing radius from the cluster center. Any underlying contamination that is still present in our `cluster stars' sample will be most pronounced at the largest distances from the cluster centre. In particular, the field star density level can be identified from the onset of a uniform density value in the radial density profile with increasing radii, which is clearly seen for all 4 clusters (see Fig., \ref{fig:RP_sb}). The radius range in the density profiles used to determine the field star density was set as from the point where the flattening begins to the outermost annulus that had 100\% areal coverage within the field of view. The radial range used for each cluster was 60 -- 100 arcmin for NGC 1851 and M2, 30 -- 110 arcmin for NGC 5824 and 30 -- 55 arcmin for NGC 1261. We then subtracted these field star densities from the total densities to generate the cluster radial density profiles. 

\subsection{Radial Density Profile}\label{chap:RDP}

The resulting field-subtracted radial density profiles (along with the star counts of M2 from Paper \RNum{1}) are displayed in Fig. \ref{fig:RP}. Accompanying our star counts, we have incorporated the surface brightness data from \cite{1995AJ....109..218T} that was scaled to our data in the regions they overlap. This allowed us to have coverage into the central regions of the clusters. We have then used the code LIMEPY, a \textit{python}-based solver of distribution functions \citep{2015MNRAS.454..576G} to fit the surface density distributions. We fit \cite{1966AJ.....71...64K} and \cite{1975AJ.....80..175W} model profiles to our star counts through a least-squares method, and the parameters of the best-fit models are displayed in Table \ref{tab:LIMEPY}. While both the King and Wilson models will give similar descriptions of the central regions of the cluster, the Wilson model has a more extended profile than the King model. This is due to the addition of an extra linear term in the distribution function compared to the King model description. This generates a more extended profile for the same central potential \citep[see][]{2005ApJS..161..304M}. Fig. \ref{fig:RP} shows that the Wilson models describe the data better than the King models in the outer parts. Having said that, it is easily seen that for most of the observed profiles, the outer parts deviate from both models. We fit power laws to the profiles from the point of deviation from the models for NGC 1851 and NGC 1261. NGC 1851 sees a deviation from the Wilson model near 16.5 arcmin, and the profile declines at a rate of $\gamma=-1.5 \pm 0.2$ beyond this point. NGC 1261 declines at a sharper rate ($\gamma=-3.8\pm0.2$) beyond the apparent deviation from the Wilson model near 6.3 arcmin. These profiles suggest that there are cluster stars beyond the limiting radius of both the King and Wilson models, or stars that are "extra tidal". Both these clusters have been previously suggested to posses extra tidal stars \citep[e.g.,][]{2012MNRAS.419...14C} and NGC 1851 is definitively known to possess an extended envelope \citep{2009AJ....138.1570O}, though the overall morphology of the envelope is unknown. 


The NGC 5824 profile, however, is well fit by a large $c$-value Wilson model over most of the observed radial range.  We also find that outside of the core radius (0.08 arcmin), the observed profile is well described by a power-law with index $\gamma = -2.2 \pm 0.02$ until a radius of $\sim5.5$ arcmin where the observed profile (and the Wilson model fit) begin to curve downwards. Additionally, at the largest radii, there is tentative evidence for an upwards deviation from the model fit similar to what is seen in the other clusters.  This deviation commences at about 25 arcmin and the density points beyond this radius are described by a power-law with index $\gamma = -1.3 \pm 0.5$.  We caution, however, that this possible detection of a very extended envelope around NGC 5824 remains uncertain; a 1 sigma increase the adopted background is sufficient for it to cease to be detected.

\begin{figure*}
  \begin{center}  
    \includegraphics[]{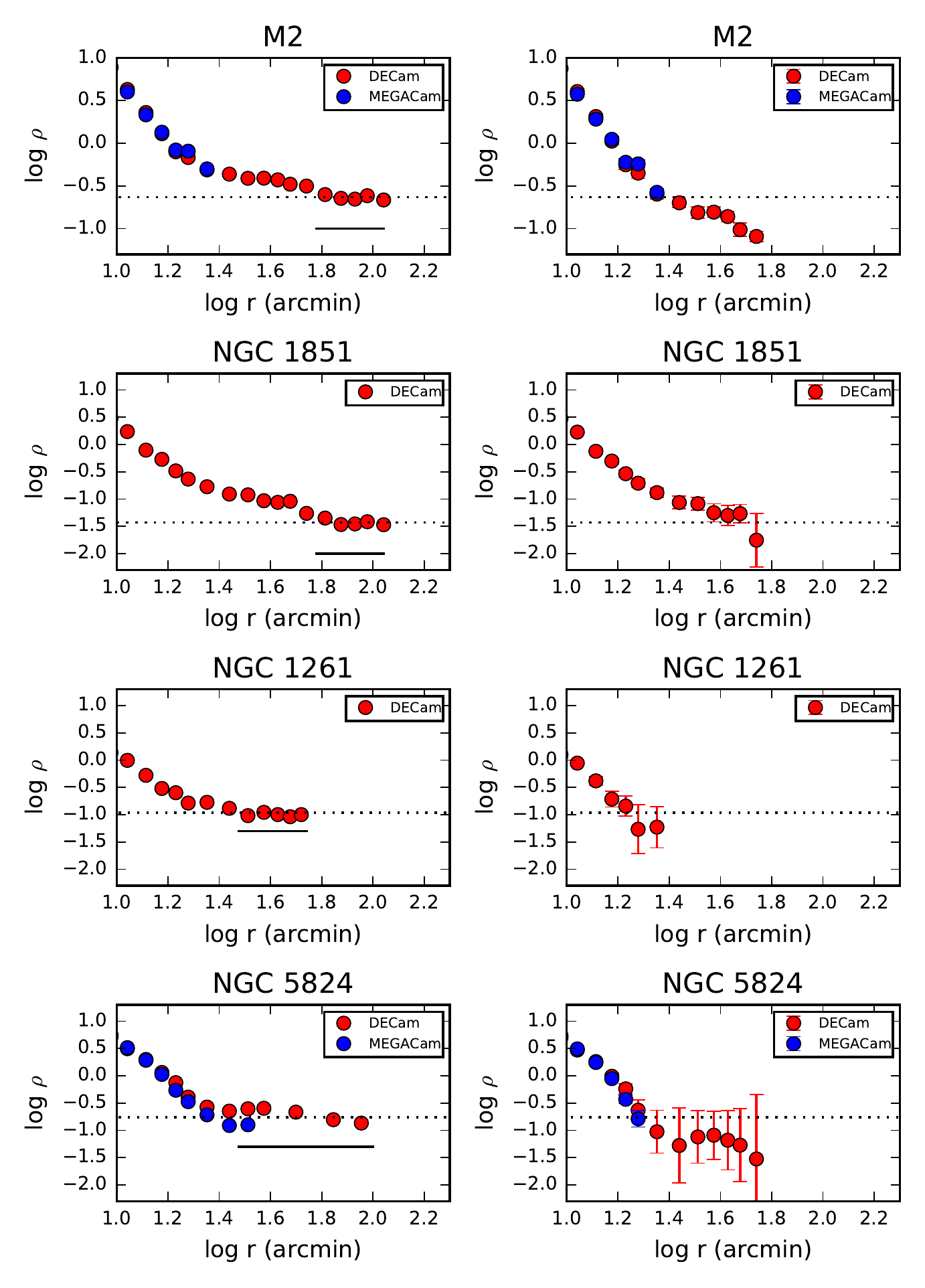}
  \end{center}
\caption{Left column: The outer radial profile of our clusters prior to field star subtraction. The underline indicates the region used to estimate the field star density level. The uncertainties here are smaller than the plot points. Right column: The outer radial profile of our clusters after field star subtraction. In both cases, the dashed dotted line indicates the determined field star density level. Full profiles and associated models are presented in Fig. \ref{fig:RP_sb}.}
\label{fig:RP_bs}
\end{figure*}

\begin{figure*}
  \begin{center}  
    \includegraphics[]{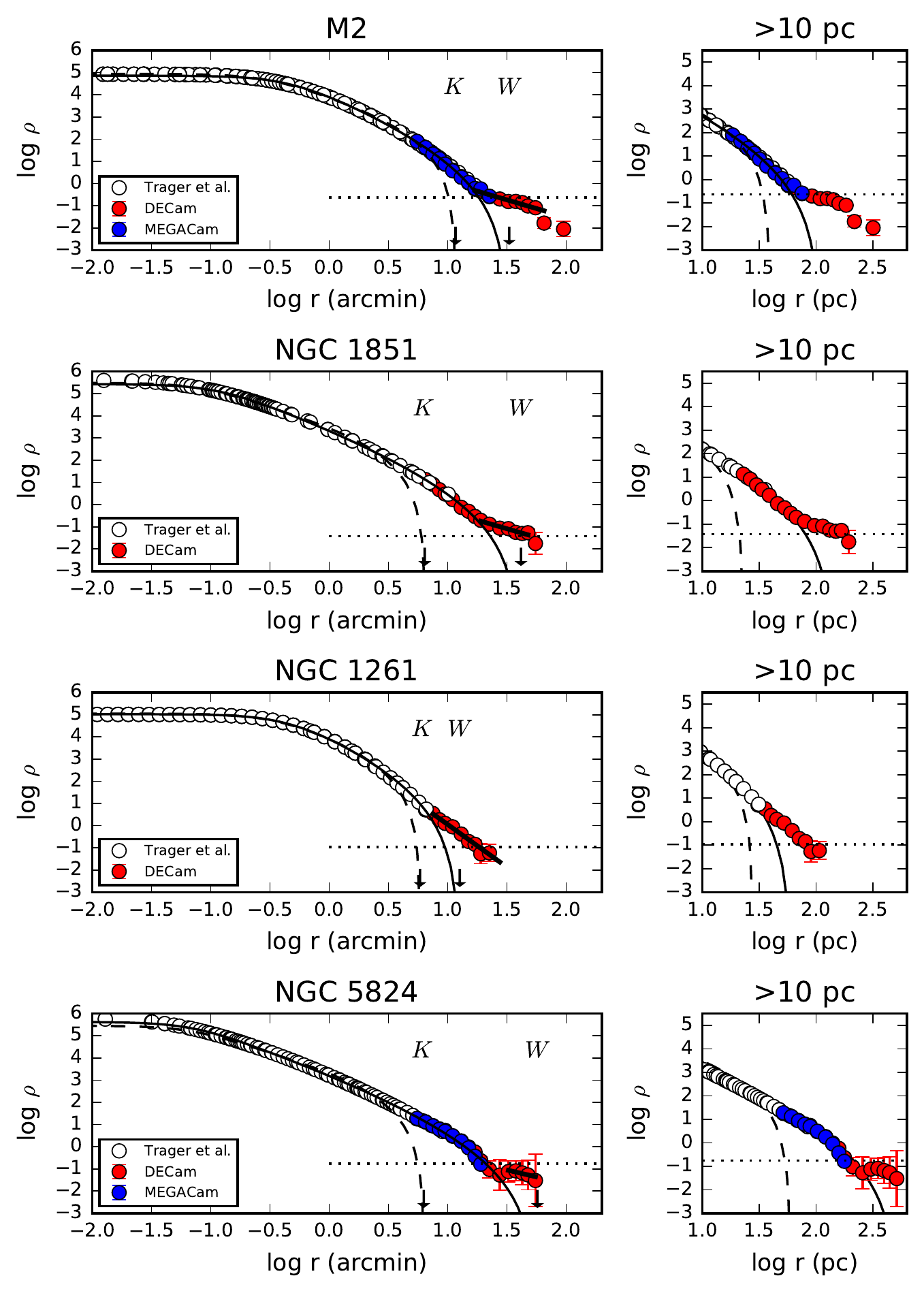}
  \end{center}
\caption{Radial density profiles for M2, NGC 1851, NGC 1261 and NGC 5824. The white points are surface photometry measurements from \citet{1995AJ....109..218T}, scaled to our star counts. The LIMEPY King (1966) and Wilson (1975) models are displayed by the dashed and solid lines respectively. Arrows pointing to the axis indicate the tidal radii of the King and Wilson models and are labelled with "K" and "W" respectively above the profiles. The horizontal dotted line shows our calculated background level. Star counts that deviate away from the Wilson (1975) models are fit with a power-law, indicated by the black bolded solid line. Left column: full profile as a function of radius in arcmin. Right: The outer regions (beyond 10 pc) of the profiles.}
\label{fig:RP}
\end{figure*}

\begin{figure*}
  \begin{center}  
    \includegraphics[width=\textwidth]{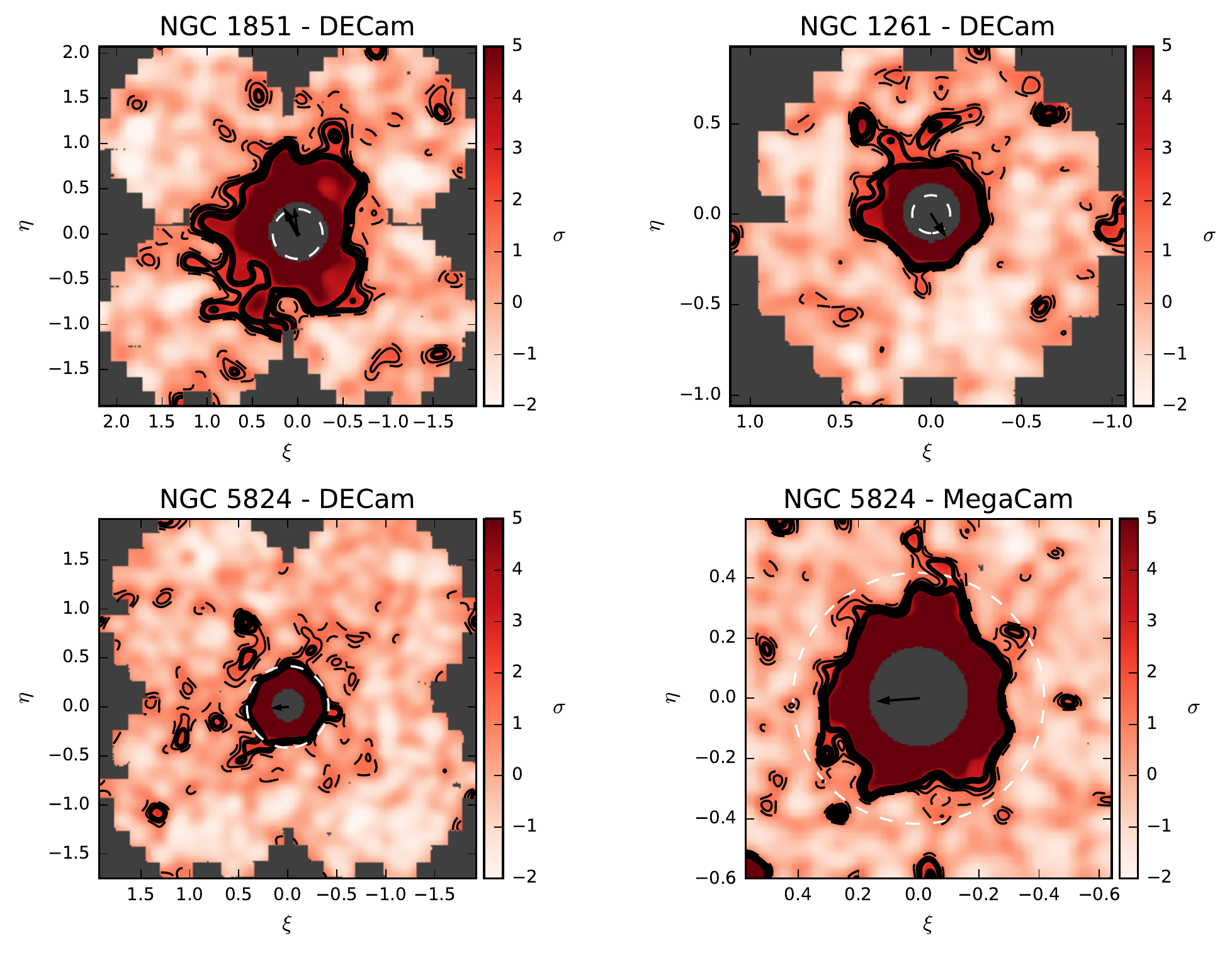}
  \end{center}
\caption{2-D density distributions from our DECam and MegaCam observations. Top row: NGC 1851 and NGC 1261. Bottom row: NGC 5824 DECam and MegaCam observations. Colour scheme depicts different levels of the standard deviation of the mean background density, with contours indicating 1.5, 2 and 3$\sigma$ and thicker contours implying higher significance. We have excluded in the inner regions of the clusters to enhance the clarity of the envelopes. Radius of areas excluded: $20\arcmin$ (NGC 1851), $10\arcmin$ (NGC 1261 and NGC 5824). Also plotted are arrows located at each clusters' centre indicating the direction to the Galactic center, and a bold arrow indicating proper-motion if it has been measured \citep[e.g.][]{1999AJ....117.1792D,2006ApJ...652.1150A}. The white ring indicates the radius of departure from the Wilson (1975) models for NGC 1851 and NGC 1261 (16.5 and 6.3 arcmin, respectively) and the 3$\sigma$ detection radius for NGC 5824 (24.5 arcmin).}
\label{fig:2dd}
\end{figure*}

\subsection{Field Subtraction and 2D Density Distribution}\label{sec:2DBS}
The radial density profiles for all our clusters show a departure from what is predicted by the best-fit LIMEPY models. The 2-dimensional (2D) surface density distribution will be able to show exactly how the extra tidal stars are distributed and whether there are any streams, tails or envelopes in the fields of view. The techniques for finding the 2D field-subtracted distribution have been improved from Paper \RNum{1}, and are similar to the techniques used in \cite{2016MNRAS.460...30R}. Specifically, we have updated how the `field' contamination was identified and removed. We created smoothed distributions of both catalogs (smoothing parameters and bin widths for each cluster are presented in Table \ref{tab:param_detect}), in their respective gnomonic coordinates transformed from R.A. and DEC (J2000). In order to remove the inevitable field contamination in our `cluster' distribution, we first created a flat-field out of the `field' distribution, by scaling the mean `field' bin density to one. The `cluster' distribution was then divided by the flat field. The original `field' distribution was normalized to the `cluster' distribution by scaling the densities to the outer regions in the `cluster' distribution, where there is no evidence of cluster populations. The normalized `field' was subsequently subtracted from the `cluster' distribution. This created our field-subtracted 2D density distribution of cluster stars. A bin was deemed as significant if its value (smoothed surface density of cluster stars) was more than three standard deviations (denoted as $\sigma$ where $\sigma$ represents one standard deviation) above the mean surface density value in the background subtracted 2D distribution. This process was completed for each cluster individually, per camera. To search for any cluster-related over-densities in the distributions, the cluster and its immediate periphery were masked as so they do not influence the statistics. This region was determined from their non background-subtracted radial profiles, and is presented in Table \ref{tab:param_detect}. The final 2D distributions for the three clusters are displayed in Fig. \ref{fig:2dd}. To help contrast the low surface brightness features in the 2D maps, we have excluded the most central regions of the clusters where the surface density values are many multiples of $\sigma$.

We will discuss the 2D distributions in the following section, but before we do, we explored the reality of a number of 2$\sigma$ detections (hereafter over-densities) that do not appear to be spatially connected to the cluster envelopes. Such features were also seen in 2D distribution of M2 (see section 3.1.4 in Paper \RNum{1}). We explored the probability of these over-densities being either somehow connected to the cluster+envelope systems, or just fluctuations in the background distributions. As described in Paper \RNum{1}, we defined a parameter $\zeta$ that was determined by exploring the number of cluster stars in any given overdensity with respect to a series of random sampling of cluster stars from the complete stellar catalog through a series of Monte Carlo simulations as described in Paper \RNum{1}. The simulations were conducted for each over-density beyond the central cluster+envelope detection for each cluster in the DECam imaging. No over-densities for any cluster detected beyond the cluster+envelope system appear to be significant (i.e. have $\zeta > 3$). 

\begin{table}
\begin{center}
\caption{Parameters used to calculate the 2D density maps.}
\label{tab:param_detect}
\begin{tabular}{@{}ccccc}
\hline \hline
Cluster&Camera&Bin&Smoothing&Masked\\
&&Width&&Region$^{a}$\\
\hline
NGC 1851&DECam&1.2$\arcmin$ x $1.2\arcmin$&$6\arcmin$&$\leq 60\arcmin$\\
NGC 1261&DECam&0.6$\arcmin$ x 0.6$\arcmin$&3$\arcmin$&$\leq 20\arcmin$\\
NGC 5824&DECam&0.6$\arcmin$ x 0.6$\arcmin$&4.8$\arcmin$&$\leq 20\arcmin$\\
&MegaCam&0.36$\arcmin$ x 0.36$\arcmin$&1.8$\arcmin$&$\leq 20\arcmin$\\
\hline
\end{tabular}
\end{center}
$^{a}$Radius masked for mean bin value calculation.
\end{table}

\section{Analysis}
\subsection{M2}
We presented the existence of a diffuse stellar envelope embedding M2 in Paper I. In Paper I, we calculated the characteristics of the envelope based on the deviation from the fit of a \cite{1962AJ.....67..471K} model. As we have developed new techniques for analysing the radial density profiles, we decided to revisit the radial profile of M2 with the improved methods discussed in section \ref{chap:RDP}. From the LIMEPY models, we found M2 deviates from both the King and Wilson profiles. Defining the stellar envelope as the deviation from the Wilson model, we find the star counts deviate from the Wilson model at approximately 17$\arcmin$, and decrease with a power law rate of $\gamma = -1.6 \pm 0.2$ beyond this point. In light of this new information, we recalculated the estimated mass ratio in the envelope by numerically integrating the observed radial density profile. The mass ratio was determined as the ratio between the integral of the profile from the point of deviation from the Wilson model to the extent of the profile (which we have defined as the envelope) with respect to the combined integral of the Wilson model (which defines the cluster) and the envelope (see Paper \RNum{1}). Under the new models, we integrated the profile between 17$\arcmin$-70$\arcmin$ and found the envelope contains $1.06 \pm 0.04\%$ of the combined mass of M2 and its envelope, where the uncertainty is estimated by varying the outer regions of the radial profile by one standard deviation. 

\subsection{NGC 1851}
\begin{figure*}
  \begin{center}  
    \includegraphics[]{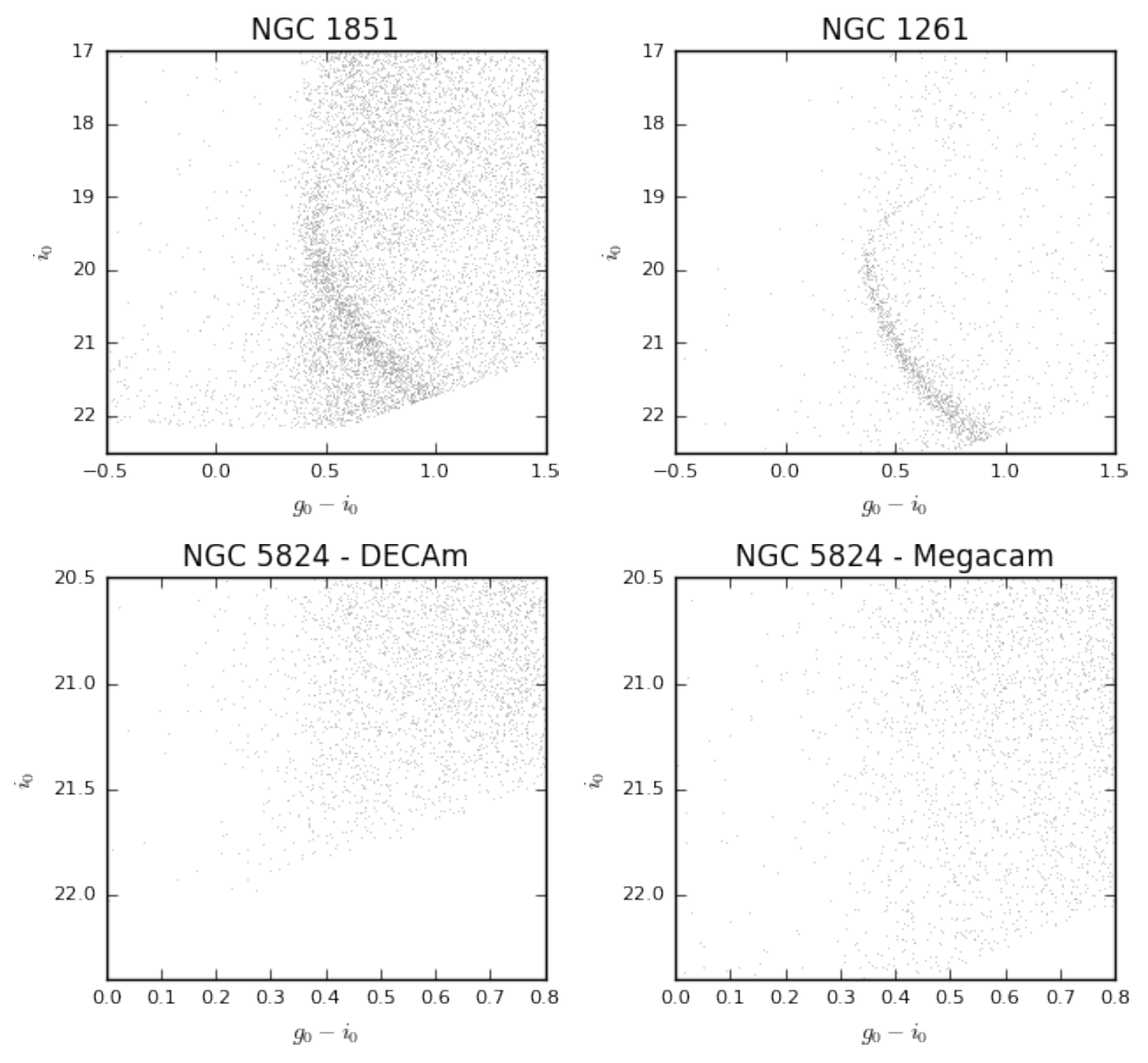}
  \end{center}
\caption{Top row: CMD of the stellar envelopes ranging from the radius of deviation from the LIMEPY models to the extent of the 3$\sigma$ detection belonging to: NGC 1851 (left, stars of radius $16.5\arcmin < r < 67.5\arcmin$) and NGC 1261 (right, stars of radius $6.3\arcmin < r < 22\arcmin$). Both CMDs show clear signs of the clusters' main sequence. Bottom row: CMD of the region surrounding NGC 5824 between the radius of deviation from the LIMEPY models to the outermost radial bin that had a measured density value that is non-zero: $25\arcmin < r < 50\arcmin$. No obvious main sequence is seen.}
\label{fig:cmd_env}
\end{figure*}

In the upper left panel of Fig. \ref{fig:2dd} the envelope of NGC 1851 is clearly visible, well beyond the excluded inner 20$\arcmin$. We see no distinct tail-like feature similar to those seen in Palomar 5 \citep[see][]{2003AJ....126.2385O} or NGC 5466 \citep{2006ApJ...637L..29B,2006ApJ...639L..17G}, nor is there an obvious large stream nearby; the structure we see is centered on NGC 1851 and does not extend across the entire field of view. The envelope we find extends out to a radius of 67.5 arcmin at the 3$\sigma$ detection level, which at a distance of 12.1 kpc from the Sun corresponds to $\sim240$ pc. This value is in good agreement with the extent of the radial profile and comparable to the $\sim250$ pc given in \citet{2009AJ....138.1570O}. We have also presented the CMD for the region beyond the nominal Wilson tidal radius in the upper left panel of Fig. \ref{fig:cmd_env} and the main sequence is clearly evident.

With the aid of the astroML python module\footnote{\url{http://www.astroml.org/}}, we fit a bivariate Gaussian to the envelopes to uncover any possible orientation or elongation. Using ``cluster stars" between the radial distance of the deviation from the Wilson model to the limit of the 3$\sigma$ detection (16.5$\arcmin$ - 67.5$\arcmin$), we find an ellipticity $e=0.17 \pm 0.04$ with a preferred orientation of $\theta=176\degr \pm 19\degr$, though this position angle is poorly constrained. The moderate ellipticity of the envelope agrees with the central regions of the cluster (< 16.5$\arcmin$), $e=0.11 \pm 0.01$, though the position angle of the cluster ($\theta=70\degr \pm 2\degr$) does not match with the envelope.

The star counts of NGC 1851 (Fig. \ref{fig:RP}, top left) revealed a clear deviation from the King and Wilson models. Of these two models, the Wilson model better describes the star counts, but in either case, there are stars beyond the model profiles, decreasing a rate described by a power law of index $\gamma =-1.5 \pm 0.2$. \citet{2009AJ....138.1570O} found a similar relation with a best-fit power law index of $\gamma = -1.24 \pm 0.66$ in agreement with our result. Taking the envelope to extend from 16.5' to 67.5', we find that it contains $0.92 \pm 0.08\%$ of the total mass of the cluster+envelope system. This is more than the $0.1\%$ reported in \citet{2009AJ....138.1570O}, though it is similar to the ratio of M2's envelope, $\sim1.1\%$. Based on the Wilson model, NGC 1851 is highly concentrated, $c=2.55 \pm 0.03$. This is amongst the highest concentrations for Galactic globular clusters (see the Wilson models presented in \cite[][]{2005ApJS..161..304M}).

\subsection{NGC 5824}
\cite{1995AJ....109.2553G} searched for tidal tails in a large sample of Globular Clusters using photographic plates. Amongst those, NGC 5824 was one of the clusters whose radial profile suggested the presence of stars beyond the King tidal radius. We find that the over the radius 1.5 to 45 arcmin (see Fig. 6 in \citealt{1995AJ....109.2553G}) the profile follows a power law of index $\gamma = -2.2 \pm 0.1$. This is consistent with our findings if we fit a power law over our radial density profile, $\gamma = -2.20 \pm 0.02$. When compared to the data of \cite{1995AJ....109.2553G} our star counts are consistent with theirs but begin to differ for radii beyond 15 arcmin. Specifically, as seen in Fig. \ref{fig:RP}, we see a drop away from the -2.2 power law slope beyond 13 arcmin that is not present in the \cite{1995AJ....109.2553G} data. Given our superior photometric precision and our explicit allowance for the significant variable reddening, we believe our data are more reliable than those of \cite{1995AJ....109.2553G} in these outermost regions. 

We find the stars counts are reasonably well described by a Wilson model. The limiting radius from the Wilson model is $\sim 530$ pc for NGC 5824, with only a small group of clusters having a similar radius or larger (e.g., NGC 5634: $\sim537$ pc, NGC 6356: $\sim589$ pc, NGC 6139: $\sim676$ pc, See \citealt[][]{2005ApJS..161..304M}). Furthermore, NGC 5824 is found to be remarkably concentrated, $c = 2.86 \pm 0.16$, in very good agreement with the Wilson model fit in \citet{2005ApJS..161..304M}, $c=2.87 \pm 0.08$. In fact, NGC 5824 is amongst the most concentrated clusters in the Milky Way. Similarly massive clusters like NGC 5824 with a comparable concentration include NGC 5634 ($c = 2.79 \pm 0.08$) and NGC 6139 ($c = 2.95 \pm 0.08$) \citep{2005ApJS..161..304M}. However, we note that the data sets featured in \citet{2005ApJS..161..304M} do not cover a similar radial extent as our profiles. As a result, their values for limiting cluster radii and concentration indices may not be as accurate when compared to our profiles.

\begin{figure}
  \begin{center}  
   \includegraphics[width=\columnwidth]{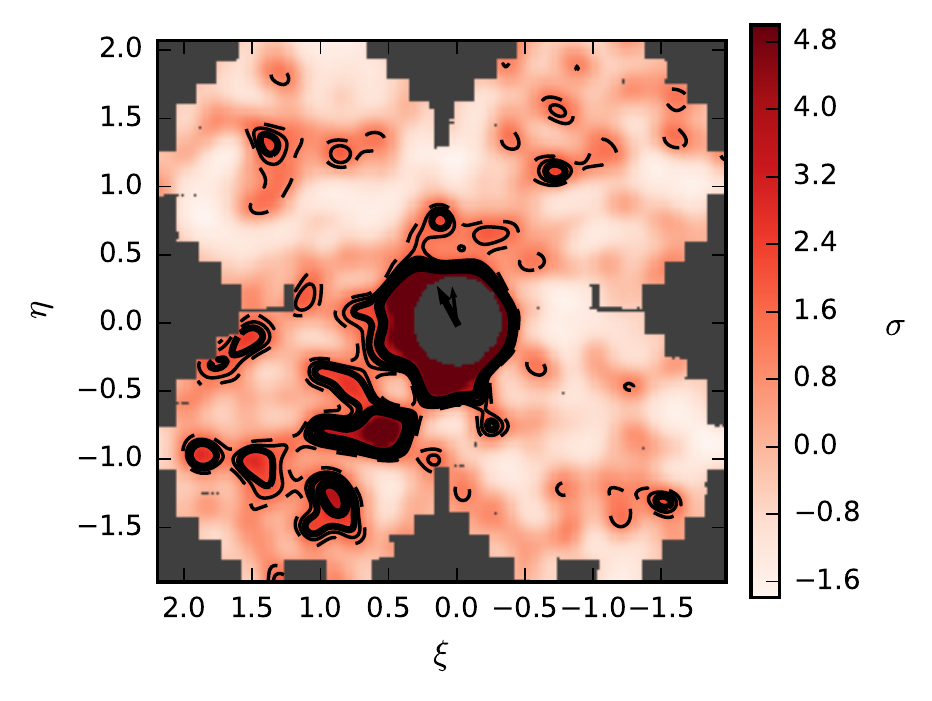}
 \end{center}
\caption{2D density distribution of NGC 1851 when the detection limit is restricted to 1.5 mag below the main sequence turn off. There is some structure beyond the envelope in this image that has no match in Fig. \ref{fig:2dd}, but none of these over densities were determined to be a real feature ($\zeta > 3$) in the 2D map. } \label{fig:de18}
\end{figure}

Both the 2D distributions from the MegaCam and DECam imaging of NGC 5824 show that the cluster does not obviously have an extended envelope, based on the interpretation of the fitted Wilson model. We find the cluster is detectable (3$\sigma$ detection) to a radius in the DECam imaging of 24.5$\arcmin$ ($\sim$230 pc). The cluster, between 5$\arcmin$ and 24.5$\arcmin$, has a low ellipticity, $0.18 \pm 0.01$, and position angle of $87\degr \pm 2\degr$. We note that in Fig. 8 there are a number of moderate significance (2.0 - 3.0 $\sigma$) detections at large radii in the DECAM imaging.  These may be signs of extended structure beyond our detection limit.

Despite the differences in scales and adopted smoothing, and the differences in the outer radial profile, we find the 2D distribution DECam observations (Fig. \ref{fig:2dd}) to be in broad agreement with the 2D distribution shown in Fig. 13 of \cite{1995AJ....109.2553G}. Specifically, we have recovered a similar looking features in the North West to North East, although we do not have detections to the South of the cluster.  There are no signs of any large stream-like structure in the field of view, or any detection of tidal tails.

We further find that beyond $\sim25\arcmin$, the star counts follow a power law out of index $\gamma = -1.3 \pm 0.5$  to a distance of $\sim50\arcmin$ (470 pc). While still within the Wilson limiting radius, the power law index is similar to both NGC 1851 and M2. These star counts are within $1\sigma$ error of the background uncertainty, and given the uncertainty on the power law index, it is ambiguous whether these star counts are describing a real feature of NGC 5824. The lower row of Fig. \ref{fig:cmd_env} shows the CMD all stars between radial region that follows the shallow power law description ($25\arcmin < r < 50\arcmin$), and there is no apparent evidence for the main sequence of NGC 5824. We decided to perform a check to see how the envelope of NGC 1851 appears if we restrict the photometric depth of our NGC 1851 data to match that of NGC 5824 (1.5 mag below the MSTO). Fig. \ref{fig:de18} shows that the majority of the envelope disappears, leaving behind a considerably smaller envelope. There are some over densities present in this figure that are not in Fig. \ref{fig:2dd}, but none of which have a $\zeta$ value greater than three. Based on this comparison it is conceivable that NGC 5824 could harbour a large diffuse envelope that would be revealed if substantially deeper data were available. 

We can estimate the mass ratio in the tentative outer stellar structure by employing the same techniques we have used for NGC 1851 and M2. Beyond the point of deviation of the Wilson model ($\sim25\arcmin$, $\sim235$ pc), we find the mass ratio to be $0.82 \pm 0.05\%$, similar to the mass ratio of NGC 1851's envelope. Considering the similarities between the possible extended stellar structure with what has been calculated from NGC 1851, deeper photometry may uncover a substantial number of extra-tidal stars belonging to NGC 5824.

\subsection{NGC 1261}
DECam imaging of NGC 1261 reveals the existence of a small, but detectable, envelope. As for NGC 1851 and NGC 5824, we see no evidence for any 2-arm axisymmetry, with the envelope detectable out to 22$\arcmin$ ($\sim$ 105 pc). The debris appears symmetric with an ellipticity of $e = 0.04 \pm 0.01$ and an associated position angle of $\theta = 79\degr \pm 9\degr$ East of North between 6.3$\arcmin$-22$\arcmin$, the apparent radial extent of the envelope. The envelope is less massive than previous envelopes this study has uncovered; the low-surface brightness feature contains $0.42 \pm 0.03\%$ of the total mass of the NGC 1261 system. Further, the upper left panel of Fig. \ref{fig:cmd_env} clearly displays that the stellar population of NGC 1261 is detected beyond the nominal Wilson tidal radius.

Compared to the other clusters in this study, the star counts for NGC 1261 drop off at a steeper rate with $\gamma = -3.8 \pm 0.2$ instead of $-2 < \gamma < -1$.  Moreover, the power law outer profile fit for NGC 1261 found by \cite{2012MNRAS.419...14C}, $\gamma = -3.68^{+0.07}_{-0.17}$ is consistent with our findings. The envelope is detectable out to a less than half the distance of the envelopes surrounding M2 and NGC 1851 (see also right column of Fig. \ref{fig:RP}) and the extent of NGC 5824 itself. Combining this difference in power law slope with the fact that the NGC 1261 envelope contains notably less mass, this then suggests that the NGC 1261 envelope may have an origin that is different from that for the other envelopes uncovered in this study.

\section{Discussion}
\subsection{Origin of the envelopes}
This study has presented evidence for the existence of diffuse extended stellar envelopes surrounding four massive Galactic globular clusters. The envelopes extend beyond the King and, in most cases, Wilson model fit limiting radii. M2 and NGC 1851 show a well-defined break from the best fit Wilson model, which is already substantially more extended than the best fit King model. Beyond the Wilson model, there is a power law distribution of $\gamma = -1.6 \pm 0.2$ for M2 and $\gamma = -1.5 \pm 0.2$ for NGC 1851. The envelopes extend to at least 240 pc in radius in both cases and contain approximately $1.1\%$ and $0.9\%$ of the system mass. NGC 5824 is well fit by a very extended Wilson model ($r_t = 533 \pm 7 $pc). We note that, despite no definitive detection of a diffuse stellar envelope, the apparent size of NGC 5824 is larger than that for the M2 and NGC 1851 envelopes. NGC 1261 is like M2 and NGC 1851 in that there is a well defined break from the best fit model, but the envelope is detected to much smaller radii ($\sim105$ pc), the power-law slope is substantially steeper ($\gamma = -3.8 \pm 0.2$) and the fractional mass in the envelope is less ($0.4\%$). A summary of these results are in Table \ref{tab:param_envelopes}.

It is natural to ponder how these envelopes came to be; whether they are born out of the dynamical evolution of globular clusters in the Milky Way, or are perhaps linked to the remains of dwarf galaxies that the Milky Way accreted some time ago. We proposed in Paper \RNum{1} that based on: (a) orbital information, (b) the distribution and shape of the detected extended envelope, (c) the presence of internal Fe abundance variations, and (d) the existence of peculiar stellar populations in the cluster CMD, the M2 cluster+envelope system may have its origins in a long since accreted dwarf galaxy, rather than being a natural product of the dynamical evolution of globular clusters. We will now explore these two scenarios with respect to NGC 1851, NGC 5824 and NGC 1261. 

Even though they spend a majority of their lifetimes away from the disk, Galactic halo globular clusters can still suffer from disruption due to tidal effects from the Milky Way. These interactions can add enough energy to the cluster for stars to escape the cluster. Modelling of this process shows that the escaped stars form long streams extending out from the cluster itself, one leading and the other trailing the cluster; otherwise known as tidal tails \citep{2006MNRAS.367..646L,2010MNRAS.401..105K}. All four clusters do not show the long, thin tidal tail structures which are seen around a small number of other clusters (e.g., Palomar 5; \citealt{2001ApJ...548L.165O,2006ApJ...641L..37G}, NGC 5466; \citealt{2006ApJ...637L..29B}). 


When a cluster passes through the disk of the Milky Way, it experiences a significant change in the gravitational potential on a short timescale. This inflicts a sudden addition of energy on the globular cluster, known as a shock \citep[e.g.,][]{1999ApJ...522..935G,2008gady.book.....B}. The loosely bound stars in the outermost regions of the cluster are much more affected by the shock than the tightly bound stars in the core. As a result, shocks contribute towards the disruption process: increasing the number of stars that can potentially escape the cluster. However, after a cluster experiences a shock, \cite{2010MNRAS.401..105K} shows that excited stars do not leave the cluster immediately. Instead, the excited stars start to populate the outer regions of the cluster within the Jacobi radii. It may take many dynamical times for the stars to escape from the cluster, leaving through the Lagrange points to create the characteristic tails. In their studies, \cite{2010MNRAS.401..105K} find the Jacobi radius of the non-core collapse clusters can be comparable to the observed tidal radius, with the ratio between to ranging from 0.8 to 1.2. Beyond this radius, \cite{2010MNRAS.401..105K} state that the surface density profiles can decrease at rate that follows a power-law like relationship with slope $\gamma \approx -4$, potentially as sharp as $-5$. This power-law can become noticeably flatter when the modelled cluster approaches apogalacticon; at this point in the clusters orbit, the power-law indices can be shallow as $\gamma \approx -1$. Other studies present similar results: the models of disrupted globular clusters performed in \cite{2006MNRAS.367..646L} show that, beginning with a cluster that follows a King profile, the evolution of the cluster can develop a power-law profile with an average index of $-3.2$. The models, therefore, show that envelopes are possible from dynamical evolution, but they generally have steep power laws and are not very extended. These are the characteristics we will use to interpret our results. 

\begin{table*}
\begin{center}
\caption{Details of the clusters and their envelopes.}
\label{tab:param_envelopes}
\begin{tabular}{@{}ccccc}

\hline \hline
Cluster&Limiting Radius&$3 \sigma$ Size Detection &Power Law Index &Mass\\
&(pc)$^{a}$&(pc)$^{b}$& ($\gamma$)&Ratio\\
\hline
M2 &$111 \pm 1$& $\sim210$&$-1.6 \pm 0.2$&$1.06 \pm 0.05 \%$\\
NGC 1261&$60 \pm 5$&$\sim105$ &$-3.8 \pm 0.2$&$0.42 \pm 0.03\%$\\
NGC 1851&$146 \pm 2$&$\sim240$&$-1.5 \pm 0.2$&$0.92 \pm 0.08\%$\\
NGC 5824$^{c}$&$533 \pm 7$&$\sim230$&($-1.3 \pm 0.5$)&($0.82 \pm 0.05\%$)\\
\hline

\end{tabular}
\end{center}
$^{a}$ Limiting radius from the Wilson Model. $^{b}$ Size of the 3 $\sigma$ detection from the 2D distribution.\\
$^{c}$ Bracketed values are properties of the tentative envelope estimated by the radial density profile.
\end{table*}

\subsubsection{NGC 1851}

NGC 1851 has been known to be embedded in a envelope since the photometric findings of \citet{2009AJ....138.1570O}. \citet{2014MNRAS.445.2971C} also found evidence for extended structure surrounding NGC 1851. However, it was unclear from those studies what the morphology of the envelope was, and whether it extended into, or was part of, a larger stellar stream. Our results are shown in the upper left panel of Fig. \ref{fig:2dd}. The envelope is clearly visible, extending well beyond the excluded 20$\arcmin$. We see that NGC 1851, perhaps because it is not near apogalacticon \citep[e.g.,][]{1999AJ....117.1792D,2006ApJ...652.1150A}, does not have a power law outer density profile consistent with the predictions of the \cite{2010MNRAS.401..105K} models. Orbital estimates for NGC 1851 suggest that the cluster has made $\sim40$ disk passages over a Hubble time, taking 580 - 685 megayears to complete an orbit of our Galaxy \citep{1999AJ....117.1792D,2006ApJ...652.1150A}. The evaporation rates calculated in \cite{1997ApJ...474..223G} show that destruction through evaporation is on a similar time scale to destruction through bulge and disc shocks, implying that NGC 1851 is not greatly susceptible to shocks (see also \citealt{1999AJ....117.1792D,2006ApJ...652.1150A}). 

NGC 1851 has been suggested before as being the remnant of a dwarf galaxy. After the discovery of the envelope by \cite{2009AJ....138.1570O}, \cite{2012MNRAS.419.2063B} modelled the cluster+envelope system to explore the formation of the system from the accretion of nucleated dwarf galaxy. Some of our results agree with what was presented in those models. The projected radial density in Fig. 5 of \cite{2012MNRAS.419.2063B}, shows the debris follows a power law slope of $\sim-2$ within a radial distances of $\leq80$ pc, becoming steeper beyond $80$ pc as $\sim-2.5$. Our observed profile ($\gamma=-1.5 \pm 0.2$) is flatter than the models, but, within uncertainties, is consistent with \cite{2012MNRAS.419.2063B} findings. The peculiar stellar populations that NGC 1851 contains also supports an origin in a dwarf galaxy. As discussed in the Introduction, the properties of the stellar populations of NGC 1851 have a lot in common with those of $\omega$ Cen, M54 and M2, all clusters for which an origin in an accreted dwarf galaxy has been postulated. Interestingly, \cite{2014MNRAS.442.3044M} found that stars in the envelope show the same Fe-spread and similar heavy element abundances as one of the two sub-giant branch populations. Collectively, the previous discussion makes a strong argument for the envelope belonging to NGC 1851 to be the last vestiges of a dwarf galaxy.

\subsubsection{NGC 5824}
The orbit of NGC 5824 is not known. The evaporation rates of NGC 5824 are similar to that of NGC 1851, the cluster is not susceptible to disk and bulge shocks \citep{1997ApJ...474..223G}. Unlike NGC 1851, NGC 5824 is located at a greater galactocentric distance. At a distance of 25.9 kpc, NGC 5824 has the largest galactocentric distance amongst the clusters presented in this study (M2: 10.4 kpc, NGC 1851: 16.6 kpc, NGC 1261: 18.1 kpc; \cite{1996yCat.7195....0H} 2010 edition). Therefore it is possible that NGC 5824 could hold onto a diffuse stellar envelope if the orbit does not take it relatively close to the Galactic center. While we do not find definitive evidence for stellar envelope, NGC 5824 is still very extended. We found NGC 5824 detected out to a radius of approximately 230 pc. As well as being similar in size to the envelopes belonging to M2 and NGC 1851, it is also comparable to the half-light radii of Local Group dwarf galaxies \citep[see][]{2009Natur.461...66M}. Furthermore, the Wilson model fit to NGC 5824 gives a truncation radius much larger than most, if not all, Milky Way Globular Clusters \citep[see][]{2005ApJS..161..304M}. The concentration parameters calculated in both the King and Wilson models (see Table \ref{tab:LIMEPY}) present more similarities between NGC 5824 and NGC 1851. 

Further, NGC 5824 also has common properties with other anomalous globular clusters such as $\omega$ Cen and M54. It is amongst the brightest clusters in the Galactic halo ($M_v = -8.83$ mag), and is second only to M54 at galactocentric distances beyond $\sim20$ kpc. NGC 5824 was reported in \cite{2014MNRAS.438.3507D} as having an internal Fe abundance variation, although \cite{2016MNRAS.455.2417R} was unable to confirm the variations, however the authors did find a star with notably different s-process abundances amongst their sample. Our observations do not provide concrete evidence for the existence of a diffuse stellar envelope surrounding NGC 5824. However, we have determined that NGC 5824 itself has a limiting radius of approximately 500pc according to Wilson model fit, much larger than the envelopes of M2 and NGC 1851. This fact and the above discussion does present encouraging results that warrant deeper photometric observations of NGC 5824 and its periphery.


\subsubsection{NGC 1261}
The final cluster presented in this paper, NGC 1261, does not appear to be similar to the others. It is not as massive as the others studied: it is approximately $60\%$ less massive than NGC 1851 for the same mass-to-light ratio, even more so when compared to the other clusters in this study. The stellar populations of NGC 1261 have not been extensively studied as for the other clusters, though \cite{2016MNRAS.tmp.1516M} do detect evidence for a possible Fe-variation in their chromosome maps. The radial profile uncovered an envelope, though it is different to the other envelopes we have discovered. It contains relatively less mass in the envelope ($\sim$ 0.4 \% compared to the 1.1\% and 0.9\% for M2 and NGC 1851) and the radial profile follows a much sharper power law, $\gamma = -3.8 \pm 0.2$, which is in good agreement with the profile fit by \cite{2012MNRAS.419...14C}. The radial profile is also consistent with the globular cluster disruption models of \cite{2006MNRAS.367..646L} and \cite{2010MNRAS.401..105K}. 


NGC 1261 does not have a known orbit. However, \cite{2014MNRAS.442.1569W} placed constraints on the orbit through the Galactocentric distance and the slope of the mass function of NGC 1261. The authors report that NGC 1261 is likely near apogalacticon, with a highly eccentricity ($e$ > 0.7) orbit.  If NGC 1261 is near apogalacticon, the debris still is compatible with \cite{2010MNRAS.401..105K} simulations. Interestingly, the destruction rate of NGC 1261 appears to not be sensitive to shocks as well, with the evaporation rate remaining constant across the \cite{1997ApJ...474..223G} models. However, the relatively low concentration value compared to NGC 1851, NGC 5824 and M2 (across both King and Wilson models) may suggest it is more susceptible to dynamical effects. We suggest that NGC 1261 and its envelope are unlike NGC 1851 and M2: its envelope appears consistent with an origin in the dynamical evolution of the cluster.

\subsubsection{Connections to Dwarf Galaxies}
Combining semi-analytic modelling of galaxy formation and the Millennium II simulation, \cite{2014MNRAS.444.3670P} explored the contributions of dwarf galaxy nuclei to GC populations in galaxies. In their study the authors described a nucleated dwarf galaxy as a GC that possesses an internal heavy element abundance spread and/or a variance in age \citep{2014MNRAS.444.3670P}. In Paper \RNum{1}, we suggested that the GC M2 met this criterion and noted that the existence of an extended stellar envelope around the cluster could be added as further evidence favouring this interpretation. The similarities between M2 and NGC 1851, particularly as regards the existence of extended stellar envelopes, and potentially NGC 5824 suggest that this interpretation could be applied to these clusters as well.  However, radial profiles like that of NGC 1261 are not uncommon: the survey completed by \cite{2014MNRAS.445.2971C} shows that in their sample many GCs have outer profiles that can be described by power laws similar to what we found for NGC 1261.  There is therefore no reason to postulate that the envelope surrounding NGC 1261 is in any way related to the remnant of an accreted dwarf galaxy.

\section{Conclusion}
We have presented the results of wide-field imaging, using the mosaic cameras MegaCam and DECam, of the outer halo globular clusters, NGC 1261, NGC 1851 and NGC 5824. Identifying clusters stars though the observed colour-magnitude diagram, we have determined that all three clusters have extra tidal stars, lying beyond the predicted limiting radius of surface brightness profiles models. NGC 1851 is found to possess an envelope $\sim240$ pc in size that contains $\sim 0.9\%$ of the system mass and is described by a power law of index  $\gamma =-1.5 \pm 0.2$. NGC 1261 is also found to be embedded in a stellar envelope, $\sim105$ pc in size and contains $\sim 0.4\%$ of the total mass of the system. The density profile of the envelope is fit with a power law of index $\gamma = -3.8 \pm 0.2$. NGC 5824 does not have a detectable envelope, though it is found to extend out to a distance of $\sim 230$ pc, which is comparable to the envelopes found around NGC 1851 and M2 from our previous study.

Some fundamental properties, such as the kinematics and element abundances, of these stellar envelopes are still unknown. With respect to disrupting globular clusters, it is unclear whether the process of heating/evaporation (either through two-body relaxation or tidal/disk shocks) can create an envelope of the size of those we see for NGC 1851 and M2. The envelopes of NGC 1851 and M2, and the overall size of NGC 5824 are all similar in size to local dwarf galaxies, and the clusters themselves have properties similar to those of M54 and $\omega$ Cen. We follow, then, to the same conclusion that these clusters could be the nucleated cores of former dwarf galaxies. The envelope surrounding NGC 1261 is consistent with those seen in dynamical models \cite[e.g.,][]{2006MNRAS.367..646L,2010MNRAS.401..105K}, favouring dynamical evolution as the likely origin.

Our results so far suggest that faint envelopes are a common feature in outer halo globular clusters. However, it is important to distinguish the differences between the envelopes we have found. While we find massive, low surface brightness envelopes surrounding already anomalous clusters, the envelope embedding NGC 1261 is different, in both relative size and luminosity. It is of interest to see if this feature is common around other more `classic' globular clusters, and whether the frequency of the envelopes are comparable to the amount of clusters with tidal tails; or indeed whether the envelopes are somehow related to the formation of tidal tails. It is obvious that more data is needed before we can start to draw connections between these two seemingly different kinds of outer envelope structure in globular clusters. Targeting clusters of similar magnitude and size such as M3, NGC 2808 or NGC 7078 and deeper imaging of NGC 5824 would be beneficial towards understanding the frequency of large stellar envelopes in massive Milky Way globular clusters and their connections to the build up of the Milky Way and its Halo.

\section{Acknowledgments}
This research is supported by an Australian Government Research Training Program (RTP) Scholarship. GDC and ADM are grateful for support from the Australian Research Council (Discovery Projects DP120101237 and DP150103294). ADM is also grateful for support from an ARC Future Fellowship (FT160100206). We thank Mark Gieles for helpful discussions and allowing use of the LIMEPY modelling codes. We acknowledge the AAO International Telescopes Support Office for travel funds for observations. We thank the anonymous referee for their comments for they have improved the quality of this paper.

This research has made use of the APASS database, located at the AAVSO web site. Funding for APASS has been provided by the Robert Martin Ayers Sciences Fund.

This paper includes data gathered with the 6.5 meter Magellan Telescopes located at Las Campanas Observatory, Chile. Australian access to the Magellan Telescopes was supported through the National Collaborative Research Infrastructure Strategy of the Australian Federal Government.

 This project used data obtained with the Dark Energy Camera (DECam), which was constructed by the Dark Energy Survey (DES) collaboration. Funding for the DES Projects has been provided by 
the U.S. Department of Energy, 
the U.S. National Science Foundation, 
the Ministry of Science and Education of Spain, 
the Science and Technology Facilities Council of the United Kingdom, 
the Higher Education Funding Council for England, 
the National Center for Supercomputing Applications at the University of Illinois at Urbana-Champaign, 
the Kavli Institute of Cosmological Physics at the University of Chicago, 
the Center for Cosmology and Astro-Particle Physics at the Ohio State University, 
the Mitchell Institute for Fundamental Physics and Astronomy at Texas A\&M University, 
Financiadora de Estudos e Projetos, Funda{\c c}{\~a}o Carlos Chagas Filho de Amparo {\`a} Pesquisa do Estado do Rio de Janeiro, 
Conselho Nacional de Desenvolvimento Cient{\'i}fico e Tecnol{\'o}gico and the Minist{\'e}rio da Ci{\^e}ncia, Tecnologia e Inovac{\~a}o, 
the Deutsche Forschungsgemeinschaft, 
and the Collaborating Institutions in the Dark Energy Survey. 

The Collaborating Institutions are 
Argonne National Laboratory, 
the University of California at Santa Cruz, 
the University of Cambridge, 
Centro de Investigaciones En{\'e}rgeticas, Medioambientales y Tecnol{\'o}gicas-Madrid, 
the University of Chicago, 
University College London, 
the DES-Brazil Consortium, 
the University of Edinburgh, 
the Eidgen{\"o}ssische Technische Hoch\-schule (ETH) Z{\"u}rich, 
Fermi National Accelerator Laboratory, 
the University of Illinois at Urbana-Champaign, 
the Institut de Ci{\`e}ncies de l'Espai (IEEC/CSIC), 
the Institut de F{\'i}sica d'Altes Energies, 
Lawrence Berkeley National Laboratory, 
the Ludwig-Maximilians Universit{\"a}t M{\"u}nchen and the associated Excellence Cluster Universe, 
the University of Michigan, 
{the} National Optical Astronomy Observatory, 
the University of Nottingham, 
the Ohio State University, 
the University of Pennsylvania, 
the University of Portsmouth, 
SLAC National Accelerator Laboratory, 
Stanford University, 
the University of Sussex, 
and Texas A\&M University.

\bibliographystyle{mn2e-2}
\bibliography{mybibs}

\label{lastpage}

\end{document}